\documentclass[lettersize,journal]{IEEEtran}
\usepackage{amsmath,amsfonts}
\usepackage{algorithm}
\usepackage{array}
\usepackage[caption=false,font=normalsize,labelfont=sf,textfont=sf]{subfig}
\usepackage{textcomp}
\usepackage{stfloats}
\usepackage{url}
\usepackage{verbatim}
\usepackage{graphicx}
\usepackage{subfig}
\usepackage{cite}
\usepackage{hyperref}
\usepackage{booktabs}
\usepackage{algpseudocode}
\usepackage{multirow}
\usepackage{epstopdf}
\usepackage{color}
\usepackage[switch]{lineno}
\begin{document}
	

\title{An Attention-Assisted AI Model for Real-Time Underwater Sound Speed Estimation Leveraging Remote Sensing Sea Surface Temperature Data}

\author{Pengfei Wu, Wei Huang*,~\IEEEmembership{Member,~IEEE,} Yujie Shi, Feng Yin,~\IEEEmembership{Senior Member,~IEEE}, Hao Zhang,~\IEEEmembership{Senior Member,~IEEE} 
\thanks{Manuscript received XX XX, 2025; revised XX XX, XXXX.}
\thanks{This work was supported in part by the National Natural Science Foundation of China under Grant 42404001 and 62271459, in part by Natural Science Foundation of Shandong Province under Grant ZR2023QF128, and in part by the Fundamental Research Funds for the Central Universities, Ocean University of China under Grant 202562022.}
\thanks{Pengfei Wu, Wei Huang and Hao Zhang are with the College of Electronic Engineering, Yujie Shi is with the School of Environmental Science and Engineering, Ocean University of China, Qingdao, Shandong 266404, China (email: wupengfei@stu.ouc.edu.cn; zhanghao@ouc.edu.cn; syj\_yqsl@163.com). Feng Yin is with the Chinese University of Hong Kong (Shenzhen), Shenzhen, 518172, China (email: yinfeng@cuhk.edu.cn).}
\thanks{Corresponding author: Wei Huang (email:hw@ouc.edu.cn). Pengfei Wu, Wei Huang and Yujie Shi contributed equally to this work.}
}

\markboth{IEEE Transactions on Neural Networks and Learning Systems,~Vol.~XX, No.~XX, XX~2025}%
{XX \MakeLowercase{\textit{et al.}}: An Attention-Assisted AI Model for Real-Time Underwater Sound Speed Estimation Leveraging Remote Sensing SST Data}


\maketitle

\begin{abstract}
The estimation of underwater sound velocity distribution serves as a critical basis for facilitating effective underwater communication and precise positioning, given that variations in sound velocity influence the path of signal transmission. Conventional techniques for the direct measurement of sound velocity, as well as methods that involve the inversion of sound velocity utilizing acoustic field data, necessitate on--site data collection. This requirement not only places high demands on device deployment, but also presents challenges in achieving real-time estimation of sound velocity distribution. In order to construct a real-time sound velocity field and eliminate the need for underwater onsite data measurement operations, we propose a self-attention embedded multimodal data fusion convolutional neural network (SA-MDF-CNN) for real-time underwater sound speed profile (SSP) estimation. The proposed model seeks to elucidate the inherent relationship between remote sensing sea surface temperature (SST) data, the primary component characteristics of historical SSPs, and their spatial coordinates. This is achieved by employing CNNs and attention mechanisms to extract local and global correlations from the input data, respectively. The ultimate objective is to facilitate a rapid and precise estimation of sound velocity distribution within a specified task area. The comparative analysis demonstrates that the proposed approach achieves superior performance in terms of both accuracy and stability, exhibiting reduced error rates and enhanced resistance to disturbances when benchmarked against existing advanced techniques.
\end{abstract}


\section{Introduction}
\IEEEPARstart{U}{nderwater} sound velocity construction plays an indispensable role in many applications such as underwater positioning, navigation, timing, communication (PNTC) and target recognition due to its decisive factor in the signal propagation mode \cite{Erol2011SurveyLoc,Luo2021LocReview,Jehangir2024Loc}. Acoustic propagation speed in marine settings exhibits spatial variability due to the combined effects of temperature, salinity, and pressure \cite{Huang2021SSPInversion,Zhang2022Loc,Yu2023Loc,Liu2024Loc}. Generally, the sound velocity exhibits a stratified variation with depth, so that sound speed profiles (SSPs) are commonly employed to describe the distribution of sound velocity.

The methodology for acquiring SSPs has been a popular research topic for a long period. In traditional ways, SSPs can be directly measured using various instruments, including the conductivity--temperature--depth profiler (CTD), the expendable CTD (XCTD), and the sound velocity profiler (SVP). Alternatively, SSPs can be derived by inversion based on acoustic field data, which encompasses parameters such as signal transmission time or received signal strength \cite{Bonnel2021Geoacoustic,Bonnel2023SSP,Wu2023SSP,Huang2024SurveySSP,Feng2024SSP}. In general, instrument-based measurements can provide precise SSPs with high--depth resolution, but the measurement process is notably time-intensive. For instance, when the instrument is deployed at a regular speed of 50 meters per minute, it takes at least 80 minutes to measure the sound velocity value within a depth range of 2000 meters to complete the deployment and retrieval of the equipment \cite{Huang2024SurveySSP}. As a result, it is difficult to provide real-time sound velocity distributions for underwater communication and positioning through instrument-based measurement. Since the distribution of the sound speed can have an impact on the distribution of the acoustic field, investigators in related areas have come up with a range of approaches to invert the sound speed distribution. They make use of the on - site measured data of the sound field like the signal propagation time, aiming at speeding up the process of getting the sound speed values effectively. There are three main frameworks for inversion methods, including matching field processing (MFP) \cite{Tolstoy1991MFP}, compressive sensing (CS) \cite{Choo2018CS, Bianco2017CS}, and deep learning (DL) \cite{Huang2021SSPInversion, Huang2023Meta}. The proposed inversion technique enables fast acquisition of sound speed profiles, offering substantial temporal advantages over conventional conductivity-temperature-depth (CTD) sensors and sound velocity profilers (SVPs) in marine data collection. However, because of the interference of noise in sound field measurement, there is a loss in the accuracy of these sound velocity inversion methods. More importantly, it still takes a certain amount of time to measure the sound field data, and new requirements have been put forward for the deployment of expensive sound field measurement equipment. The implementation of SSP inversion techniques is inherently constrained by their dependence on active sonar measurements, thereby limiting their applicability in regions beyond the operational range of underwater acoustic monitoring networks.

Over the past few years, the fast development of underwater automated sensing equipment has provided increasingly rich vertical observation reference data for exploring the distribution of underwater sound velocity. To achieve fast estimation of SSPs and eliminate the need for underwater on-site operations, various methods have emerged that use empirical sound velocity distribution data or remote sensing data. \cite{Liu2024SSPPrediction,Lu2024LSTM,Cui2024DRPrecition} utilize empirical SSPs data to capture the changes in sound velocity from a time series perspective, enabling real-time estimation and prediction of sound velocity distribution in designated ocean areas. These SSP prediction methods have a single type of input data and low complexity of model construction. However, because of the low time resolution of historical samples, its accuracy performance is not good enough in estimating the sound velocity within a small-scale time range. In fact, the variation of sound velocity distribution in a small-scale spatial area is mainly reflected in the shallow water part, as it is more significantly affected by temperature changes. Marine remote sensing technology provides real-time and reliable high spatiotemporal resolution sea surface temperature (SST) data \cite{Huang2021SST}, which can offer initial sea surface condition constraints for estimating underwater sound velocity distribution. Xu et al \cite{{Xu2025SOM}} introduced an innovative self-organizing map architecture that integrates empirical acoustic propagation measurements with satellite-derived sea surface temperature observations to establish robust correlations between marine acoustic properties and surface thermal patterns, but the model mainly focuses on the local characteristics of sound velocity distribution, which not only fails to capture the influence of SST data on deep ocean sound velocity, but also fails to capture the spatial correlation of sound velocity distribution. As a result, the accuracy performance has not been significantly improved compared to prediction methods.

In order to fast and accurately estimate the SSP of a given task area without on-site underwater data measurement, we propose an interpretable self-attention-assisted multimodal data fusion convolutional neural network (SA-MDF-CNN) model in this paper to deeply capture the intrinsic relationships among empirical SSP data, remote sensing SST data, and spatial coordinates. The central concept focuses on extracting spatial patterns of regional sound velocity interdependencies via convolutional neural networks, while the application of attention mechanisms improves the model's capacity to learn complex interactions within multimodal data's global feature relationships. In an attempt to comprehensively appraise the overall effectiveness of the purportedly proposed SA-MDF-CNN model, we meticulously carried out a series of well - designed experiments by utilizing the Argo data originating from the expanse of the Pacific Ocean and the precisely measured data obtained from the territory of the South China Sea in the year 2023. The detailed results clearly indicate that the SA-MDF-CNN model has the distinct ability to accomplish a notably lower root mean square error (RMSE) value compared to various other existing spatial SSP construction methods available. The significant contributions presented in this research paper can be systematically summarized in the following specific ways:
\begin{itemize}
\item To achieve real-time estimation of SSP without on-site underwater data measurement, we propose an interpretable self-attention-assisted multimodal data fusion convolutional neural network (SA-MDF-CNN) model. We take into account the mutual influence among empirical sound velocity distribution, SST, and spatial position, and the attention module is developed to optimize the model's comprehensive feature extraction capabilities across spatial domains.

\item To assess the precision performance of the model, we carried out experiments with Argo data sourced from the Pacific area. Subsequently, we verified it via sea trials in the South China Sea during 2023. The RMSE of the two experimental outcomes outperforms other spatial SSP building methods.

\item To boost the model's interpretability, a visualization of the model parameter weights is carried out. With the escalation of training iterations, the weights of model parameters progressively concentrate on the shallow water region. This phenomenon is justifiable as, upon performing Empirical Orthogonal Function (EOF) decomposition on historical SSPs, it becomes evident that disparities in sound velocity distribution are more pronounced in shallow water. Consequently, the outcomes of weight visualization suggest that the model is capable of efficiently capturing the traits of sound velocity distribution.
\end{itemize}

The organization of this paper is delineated as follows. In Section 2, an in - depth elaboration on related works is offered. Section 3 puts forward the general framework and the descriptive details of the functional modules of the SA-MDF-CNN model. The outcomes of the experiment are reported in Section 4. Concluding statements are presented in the last section, Section 5.

\section{Related works}
The spatial structure of underwater sound velocity field plays a decisive role in regulating the transmission mechanism of underwater acoustic energy and the geometric characteristics of wave propagation. Therefore, real-time and accurate estimation of sound velocity distribution has significant value for applications based on communication and positioning technologies such as underwater PNTC systems, and target recognition systems.

The traditional way of obtaining sound velocity distribution is usually through direct measurement using shipborne CTD or SVP equipment, which has the advantage of high accuracy \cite{Kirimoto2024CTD,Luo2023CTD}. However, it comes with high economic costs and measurement time expenses.The progression of sensor technologies and networked architectures has led to the implementation of multiple subaqueous environmental monitoring frameworks, advancing scientific comprehension of oceanic processes. The methodological framework of underwater acoustic inversion was initially formalized during 1979 by Munk et al.\cite{MUNK1979Tomography,Munk1983Tomography}, introducing an innovative paradigm that leverages hydroacoustic propagation measurements for reconstructing  SSPs. Since then, there have been three mainstream frameworks for sound velocity inversion, namely MFP \cite{Tolstoy1991MFP}, CS \cite{Choo2018CS, Bianco2017CS}, and DL \cite{Huang2021SSPInversion, Huang2023Meta}. The MFP framework mainly consists of four steps. Firstly, EOF decomposition is adopted to extract the principal component features of the regional sound velocity distribution. Then, different feature combinations are generated to form candidate SSPs. Subsequently, acoustic field distributions were computationally modeled employing an acoustic ray tracing framework. The concluding phase involved cross-validation between the modeled pressure distributions and experimentally acquired field measurements to establish the optimal inversion parameters for the SSP. To accelerate the search speed of the optimal candidate SSP, Tolstoy introduced simulated annealing algorithm in the MFP, which improved the algorithm execution efficiency but resulted in suboptimal solutions \cite{Tolstoy1991MFP}. Afterwards, other heuristic algorithms were combined with MFP frameworks for SSP inversion, but they also obtained suboptimal solutions \cite{Zhang2012MFP,Zhang2016SSPAUVInv}. To further improve the efficiency of inversion algorithm execution, Bianco \cite{Bianco2017CS} and Choo \cite{Choo2018CS} proposed SSP inversion framework based on CS, respectively. Within the context of the CS framework, the correlation between the sound field distribution and the sound velocity distribution is formulated via a matrix based approach, which eliminates the search process for matching terms. However, the matrix relationship introduces linear approximation, thus sacrificing inversion accuracy.

The theoretical frameworks of deep neural networks (DNNs) have experienced accelerated progression in contemporary research, particularly in establishing intricate nonlinear dependencies across heterogeneous data domains \cite{Reichstein2019DLEarth,Jahanbakht2021DL}. In addition, long-term underwater environmental observations have accumulated a large amount of data for marine hydrological research, laying a data foundation for the application of deep learning underwater. To address the drawbacks of computational efficiency and accuracy loss in MFP and CS frameworks, an auto-encoding neural network model for feature mapping was introduced for SSP inversion in our prior research \cite{Huang2021SSPInversion}, in which the auto-encoder was created to extract deep robust features so as to reduce the impact of acoustic field measurement errors on the accuracy of sound velocity inversion. Considering the insufficient accumulation of historical sound velocity data in some ocean areas, the DL model is prone to over-fitting and reduces accuracy performance. Therefore, we introduce a meta-learning framework tailored for few-shot scenarios to enhance the rate at which the model converges \cite{Huang2023Meta}. However,the methodologies aforementioned, which are predicated on MFP, CS, and DL, all hinge on acoustic field data that is measured in real - time, thereby imposing stringent requirements on the arrangement of underwater observation equipment. Therefore, these methods not only face high equipment economic costs, but also have limited application scope for areas without sonar measurement systems.

\begin{figure*}[!htbp]
	\centering
	\includegraphics[width=0.8\linewidth]{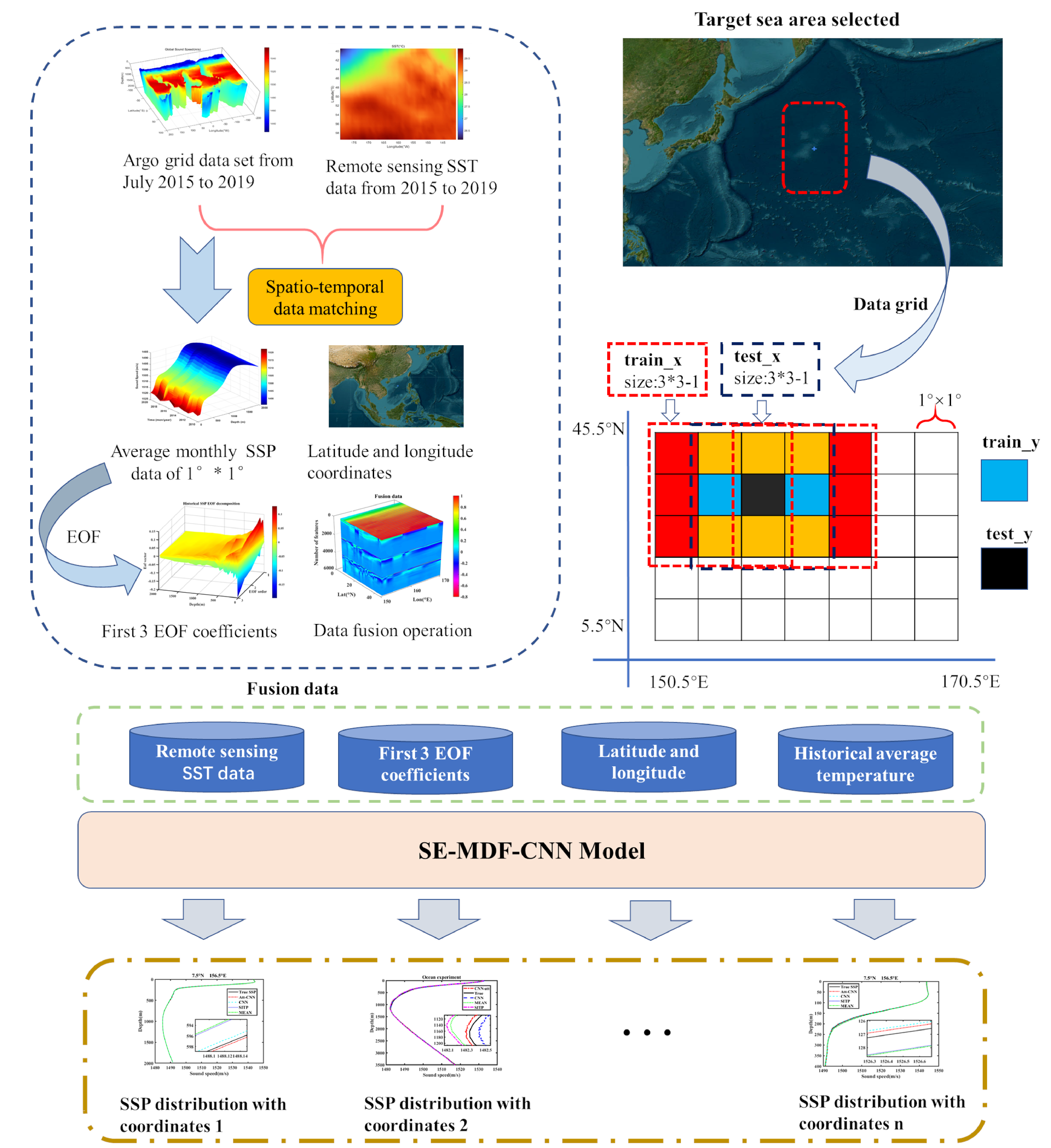}
	\caption{SSP Estimation Structure based on SA-MDF-CNN.}
	\label{fig1}
\end{figure*}

Nowadays, removing the necessity for on-site underwater data measurement has emerged as a prominent research focus in the domain of SSP inversion. In accordance with this desideratum, both Liu et al. and Lu et al. put forward a Long - Short Term Memory (LSTM) neural network model for SSP forecasting \cite{{Liu2024SSPPrediction,Lu2024LSTM}}, which solely demands empirical SSP data. Nevertheless, owing to the poor temporal resolution of prior data in the majority of marine regions, this prediction method can only describe the overall trend of sound velocity changes and is difficult to obtain high-precision SSP prediction results. In fact, the SSP estimation methods, that rely on a single data modality (sound field data or historical SSP data), are susceptible to data quality issues, such as high sound field measurement noise or low sound velocity sampling time resolution. To improve the robustness of the sound velocity estimation model, researchers have proposed some multimodal data fusion methods for estimating sound velocity that combining data from different sensors and sources to obtain more comprehensive features than a single data source. In \cite{WU2024104088} , Wu et al. achieved remarkable results in SSP invesion using a CNN network driven by data fusion. Yu et al.\cite{Yu2020RBF} proposed a radial basis function (RBF) neural network for SSP estimation that mainly using historical temperature, salinity profile data and average SSP data. Nevertheless, the model is not sensitive to varieties of sound velocity, and the estimation results often approach the average SSP profile, resulting in challenges in accurately depicting the spatiotemporal variations of sound velocity. Xu et al.\cite{Xu2025SOM} proposed a physics-inspired self-organising map (SOM) model for SSP estimation, which introduced remote sensing SST data. However, SOM is confined to capturing the impact of SST on the distribution of sound velocity at the surface, and lacks the ability to capture large-scale feature correlations. Therefore, the accuracy of sound velocity estimation is difficult to meet practical application requirements.

To obtain real-time and accurate estimation of sound velocity distribution without underwater on-site data measurement, we fully consider the historical sound velocity distribution patterns of different spatial coordinates, as well as the impact of real-time SST changes on the dynamic characteristics of sound velocity distribution, and propose an interpretable SA-MDF-CNN model. In this model, the local correlation of features will be captured through CNN and global correlation of features will be captured through attention mechanism.

\section{SA-MDF-CNN Structure for SSP Estimation}
To realize real-time and accurate estimation of sound velocity distribution, we propose an SSP estimation structure based on SA-MDF-CNN, which is shown in figure~\ref{fig1}. In this paper, the ocean area of 5.5$^\circ$N-45.5$^\circ$N and 150.5$^\circ$E-170.5$^\circ$E are selected as the research scope. The remote sensing SST data, latitude and longitude coordinates, and principal component eigenvectors derived from EOF decomposition of historical SSPs are first fused to construct the fusion data as the SA-MDF-CNN training data. Then, the multimodal fusion data of 8 grids around the task region is combined to construct the target SSP through the proposed SA-MDF-CNN model. In the following part, we will provide detailed introductions to data sources, data fusion structures, and the composition of neural network model, separately.

\subsection{Data sources}
Remote sensing SST data are provided by the National Oceanic and Atmospheric Administration (NOAA) \cite{SST} with a spatial grid resolution of 0.25$^\circ$ and a temporal resolution of 1 day. The SST data are arithmetic averaged on a monthly basis to achieve the consistency of the fused data in time sampling. The SSP data are obtained from the Chinese Observation and Research Station fo Global Ocean Argo System (Hangzhou) \cite{CAOSha-sha} with a spatial grid resolution of 1$^\circ$ and a temporal resolution of 1 month. Based on the objective analysis method of gradient-dependent correlation scale optimal interpolation, the 3D grid data of the subsurface layer (5-2000 meters) that covering 179.5$^\circ$W to 179.5$^\circ$E and 89.5$^\circ$S to 89.5$^\circ$N was constructed, and the observation profile was vertically interpolated to 58 standard layers with unequal intervals. 

\subsection{Fusion data construction}
\begin{figure}[!htbp]
    \centering
    \includegraphics[width=\linewidth]{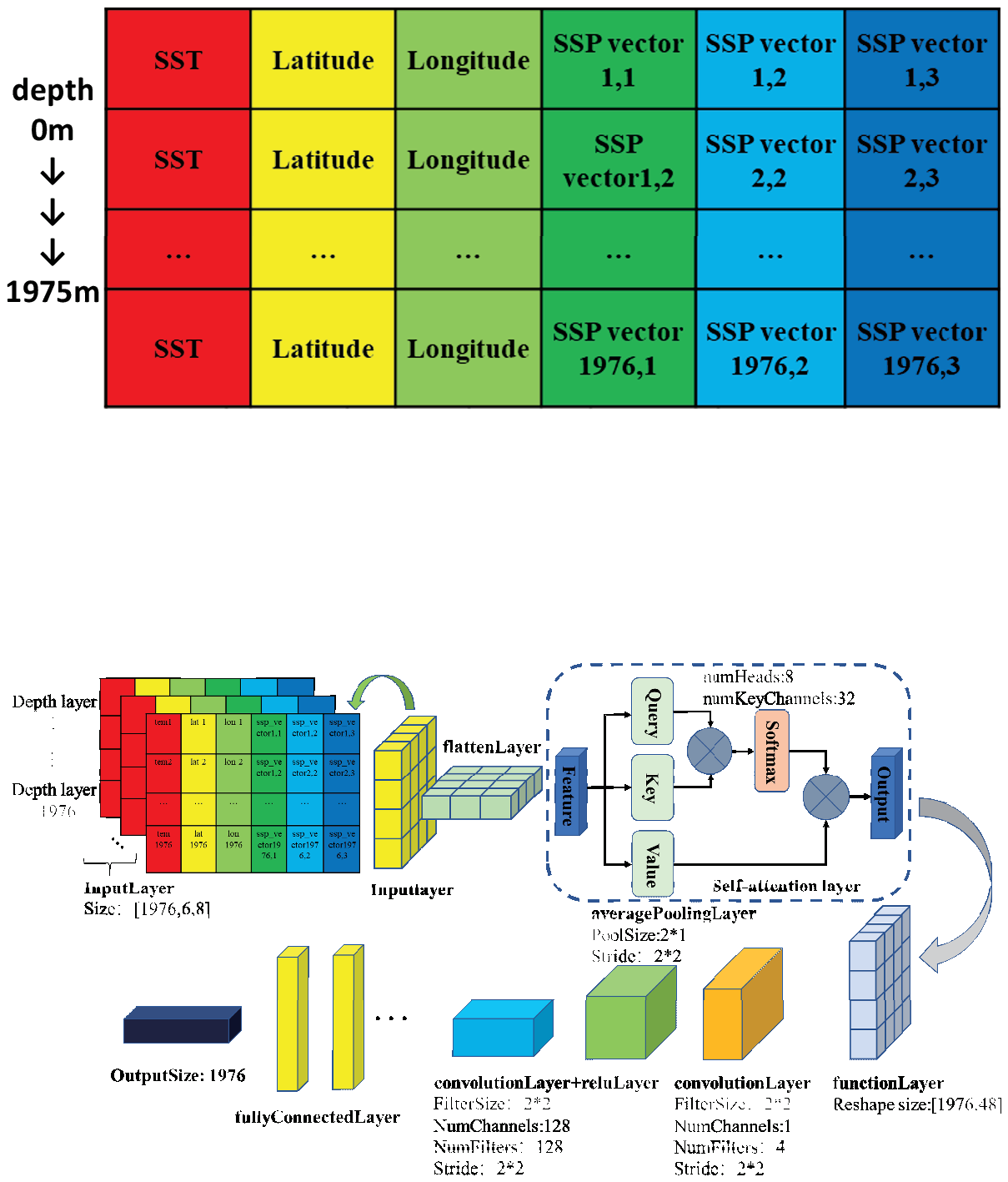}
    \caption{The data fusion structure for a single coordinate.}
    \label{fig2}
\end{figure}

\begin{figure*}[!htbp]
	\centering
	\includegraphics[width=0.9\linewidth]{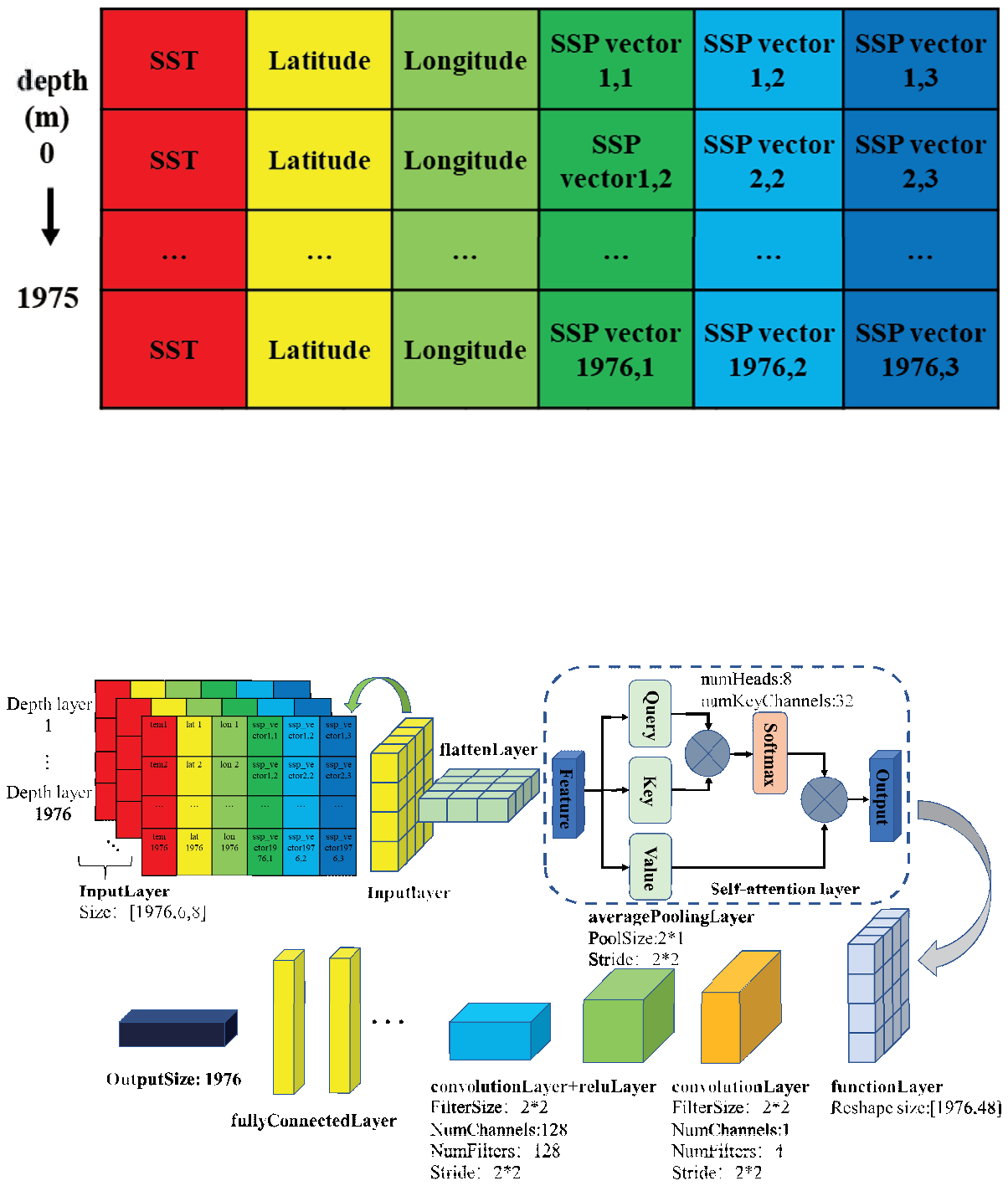}
	\caption{SA-MDF-CNN model.}
	\label{fig3}
\end{figure*}

In the selected ocean region, we grid the entire area into N longitude-scales of 1$^\circ$, M latitude-scales of 1$^\circ$, and vertical depth of H layers. For any coordinate in the grid, the grid remote sensing SST data are expressed as $T^s_{n,m}$, and the monthly average SSP data of the historical period are chosen for data fusion, with the depth of the $h$th layer expressed as $D_{n,m}^h, n\in N,m \in M$.

Within the divided $N*M$ grid ocean area, each sub-grid $\psi$ with a size of $3*3$ forms a set of training data, and each set of training data will be further divided into input data $X$ and output label data $Y$, with a sliding step of 1 for each sub grid, as shown in Figure \ref{fig1}. Specifically, for a sub-grid $\psi$, the fusion data formed by 8 surrounding coordinates will be the input data $X$, and the center SSP of each sub-grid $\psi$ will be taken as the output label $Y$. If the coordinate of the grid center is $L_Y = L_{n,m}$, then the coordinates of the input data are represented as:

\begin{equation}
\begin{aligned}
\mathcal{L_X} = [ & L_{n-1,m-1}, L_{n-1,m}, L_{n-1,m+1}, L_{n,m-1}, L_{n,m+1}, \\
& L_{n+1,m-1}, L_{n+1,m}, L_{n+1,m+1}] , \quad n \in N, \quad m \in M.
\end{aligned}
\label{eq1}
\end{equation}

In the context of a single coordinate, the configuration of data fusion is presented in figure \ref{fig2}. This configuration encompasses real-time SST data, coordinates of latitude and longitude, along with the initial three eigen-vectors of the EOF of historical SSPs. EOF decomposition, alternatively referred to as principal component analysis, is a statistical approach employed for examining the structural characteristics within matrix data and extracting the primary quantitative features of the data. By means of EOF decomposition, the variation information of the original sound velocity field can be readily compressed into the initial few principal components and their corresponding spatial functions. As a result, the essential information of the sound velocity field can be represented solely by the initial few eigen-vectors.

To obtain the EOF feature vectors, corresponding 5-year (2015 to 2019) historical SSPs from the target ocean area are first selected and then linearly interpolated with a step size of 1 m. This standardized interpolation is used to more clearly describe the differences in sound velocity values at different depths. Suppose there are $J$ SSPs, and the $j$th SSP is expressed as a vector $\boldsymbol{S_{j}}=[s_{j,1},s_{j,2},...,s_{j,h}]^T, h = 1,2,...,H, H=1976$, where $s_{j,h}$ means the sound velocity value at the $h$th depth layer of the $j$th SSP. Since the SSP has been linearly interpolated, the depth of the $h$th depth layer is h m. Then, the matrix formed by all empirical SSPs will be:

\begin{equation}
\mathcal{S_{H,J}}=\begin{bmatrix}s_{1,1}&s_{2,1}&\cdots&s_{J,1}\\s_{1,2}&s_{2,2}&\cdots&s_{J,2}\\\vdots&\vdots&\ddots&\vdots\\s_{1,h}&s_{2,h}&\cdots&s_{J,h}\end{bmatrix}.
\label{eq2}
\end{equation}

Based on equation \eqref{eq2}, the average SSP $\boldsymbol{S_{0}}=[s_{0,1},s_{0,2},...,s_{0,h}]^T$ can be calculated by averaging each row of equation \eqref{eq2}. Next, a residual matrix can be constructed by subtracting $\boldsymbol{S_{0}}$ from each column of equation \eqref{eq2}:

\begin{equation}
\mathcal{S^R_{H,J}}=[\boldsymbol{S_{1}} - \boldsymbol{S_{0}},\boldsymbol{S_{2}} - \boldsymbol{S_{0}},...,\boldsymbol{S_{J}} - \boldsymbol{S_{0}}].
\label{eq3}
\end{equation}

According to equation $\mathcal{S^R_{H,J}}$, the covariance matrix $\mathcal{C_{H,H}}$ of sound velocity can be constructed as:

\begin{equation}
\mathcal{C_{H,H}}=\frac{1}{J}\mathcal{S^R_{J,H}}*\mathcal{S^R_{J,H}}^T,
\label{eq4}
\end{equation}
where $\mathcal{C_{H,H}}$ is a matrix with orders of $H\times H$. Performing EOF decomposition on $\mathcal{C_{H,H}}$ yields eigenvectors and eigenvector coefficients:

\begin{equation}
\mathcal{C_{H,H}}\times \mathcal{E_{H,H}}=\mathcal{\lambda_{H*H}}\times \mathcal{E_{H,H}}, \label{eq5}
\end{equation}
where $\mathcal{E_{H,H}}$ is the matrix composed by eigenvectors, and $\mathcal{\lambda_{H*H}}$ is the matrix composed by eigenvalues. When sorting the eigenvector coefficients from large to small, the corresponding eigenvectors form the first few order eigenvectors.

For the $j$th SSP in $\mathcal{S_{H,J}}$, it can be recovered by:
\begin{equation}
\boldsymbol{S_{j}}=\boldsymbol{S_{0}}+\sum_{k=1}^K\alpha_k\boldsymbol{e_{k}}, \label{eq6}
\end{equation}
where $\boldsymbol{e_{k}}$ means the $k$th eigenvector, and $\alpha_k$ is the corresponding coefficient that determines the proportion of the eigenvector. When an SSP is given as $\boldsymbol{S_{tar}}=[s_{tar,1},s_{tar,2},...,s_{tar,h}]^T, h = 1,2,...,H$, the residual vector will be $\mathcal{S^R_{H,TAR}} = [\boldsymbol{S_{tar}}-\boldsymbol{S_{0}}]$. The vector of coefficients $\boldsymbol{\alpha} = [\alpha_1,\alpha_2,...,\alpha_K]^T$ can be obtained by projecting the target SSP onto the first $k$ eigenvectors of $\mathcal{E_{H,H}}$:

\begin{equation}
\boldsymbol{\alpha}=\mathcal{E_{H,K}}^T\times\mathcal{S^R_{H,TAR}}.\label{eq7}
\end{equation}

Finally, by fusing data from 8 coordinate points, input training data can be obtained. For location $L_{n-1,m-1}$, the fused data input $\mathcal{F}^{X}_{n-1,m-1}$ can be expressed as:
\begin{equation}
    \mathcal{F}^{X}_{n-1,m-1}=[\boldsymbol{T^S_{n-1,m-1}},\boldsymbol{L_{n-1,m-1}},\mathcal{E_{H,K}}],\label{eq8}
\end{equation}
where $\boldsymbol{T^S_{n-1,m-1}}=[T^s_{n-1,m-1,1},T^s_{n-1,m-1,2},...,T^s_{n-1,m-1,H}]^T$ is actually a vector composed of H copies of $T^s_{n-1,m-1}$ at location $L_{n-1,m-1}$, and $\boldsymbol{L_{n-1,m-1}}$ is similar that $\boldsymbol{L_{n-1,m-1}}=[L_{n-1,m-1,1},L_{n-1,m-1,2},...,L_{n-1,m-1,H}]^T$. The label output data is $\mathcal{S}^{Y}_{n,m}=[s^Y_{n,m,1},s^Y_{n,m,2},...,s^Y_{n,m,H}]^T$ at location $L_{n,m}$.

\subsection{SA-MDF-CNN Model}
The architectural framework of the SA-MDF-CNN is illustrated in Figure \ref{fig3}, mainly including attention module and convolutional network module. The initial layer is the input layer, the size of which is 1976*6*8, including 6 types of physical quantities (remote sensing SST, latitude, longitude, and the first 3 order eigenvectors) at 8 coordinates with a total depth of 1975 meters. Then the flattening layer serves as a flattening operation, transmuting three-dimensional fusion data into one-dimensional sequence data as input for the self-attention mechanism. The following stratum is designated as the self-attention layer, which transmutes integrated data into query, key, and value vectors via three distinct linear transformation processes. Subsequently, the inner product between every query vector and each key vector is computed, resulting in an attention score matrix. This matrix is subsequently transformed into attention weights via the softmax function. The final output features are generated through the aggregation of value vectors, facilitated by the application of attention weights. The architectural design of these layers is intended to augment the model's capacity for focusing on the inter - relationships between SSPs situated across disparate oceanic zones within the input fusion data, so as to better capture the global dependence and key information between fusion data and the sound velocity distribution in the target ocean area. 

The fourth and fifth layers are functional layers designed to reshape one-dimensional sequence data into two-dimensional data for subsequent feature extraction through convolution. The convolution layer is intended to derive additional features from the resultant data of the attention layer, which is better at capturing small-scale relationships of input features. The pooling layer serves to diminish the spatial dimensionality, thereby reducing the number of parameters and reduce computational complexity, while retaining important feature information. The penultimate and ultimate layers comprise the fully connected stratum and the regression stratum. The fully connected stratum integrates local features into global representations, subsequently passing them to the regression stratum. The root mean square error (RMSE) is designated as the loss function, which is employed to compute the gradient of the model parameters and update the model's weight parameters via the backpropagation (BP) algorithm.
\begin{equation}
Loss = \sqrt{\frac{\sum_{h=1}^{H} (\hat{s}^{Y}_{n,m,h} -s^{Y}_{n,m,h})^2}{H}},\label{eq9}
\end{equation}
where $\hat{s}^{Y}_{n,m,h}$ is the estimated sound velocity value at depth $h$ of location $L_{n,m}$, and the estimated SSP can be expressed as $\hat{\mathcal{S}}^{Y}_{n,m}=[\hat{s}^Y_{n,m,1},\hat{s}^Y_{n,m,2},...,\hat{s}^Y_{n,m,H}]^T$. The SSP estimation algorithm based on SA-MDF-CNN is provided in Algorithm~\ref{algorithm1}.

\begin{algorithm}
\caption{The SSP estimation algorithm based on SA-MDF-CNN.}
\begin{algorithmic}[1]
\Require Historical average SSP data $\mathcal{S}_{\mathcal{L_X}}^{mon}$, grid remote sensing SST $\mathcal{T}^s_{\mathcal{L_X}}$, position $\mathcal{L_X}$;
\State Initialize the parameters $\theta$, learning rate $\gamma$, and set the network parameters according to Table \ref{table1};
\For{$k = 1$ to $MaxEpoch$}
    \For{$t = 1$ to $maxBatchSize$}
        \State Select a fusion data sample \\  \quad\quad\quad\quad $\mathcal{F}^{X}_{n-1,m-1}=[\boldsymbol{T^S_{n-1,m-1}},\boldsymbol{L_{n-1,m-1}},\mathcal{E_{H,K}}]$;
        \State $\mathcal{S}^{Y}_{n,m}=[s^Y_{n,m,1},s^Y_{n,m,2},...,s^Y_{n,m,H}]^T$;
        \State $Loss_t = \sqrt{\frac{\sum_{h=1}^{H} (\hat{s}^{Y}_{n,m,h} -s^{Y}_{n,m,h})^2}{H}}$;
        \State $\hat{\theta} \leftarrow \text{Adam}(\nabla_{\theta} \frac{1}{BatchSize} \sum Loss_t, \theta, \gamma) $;
    \EndFor
\EndFor
\State Complete training and save SA-MDF-CNN model;
\State Estimate and output target SSP $\mathcal{S}_{n,m}^{es}$ at intermediate coordinates $L^{test}_{n,m}$. 
\end{algorithmic}\label{algorithm1}
\end{algorithm}

\subsection{Multi-head self-attention module}
\begin{figure*}[!htbp]
	\centering
	\includegraphics[width=0.8\linewidth]{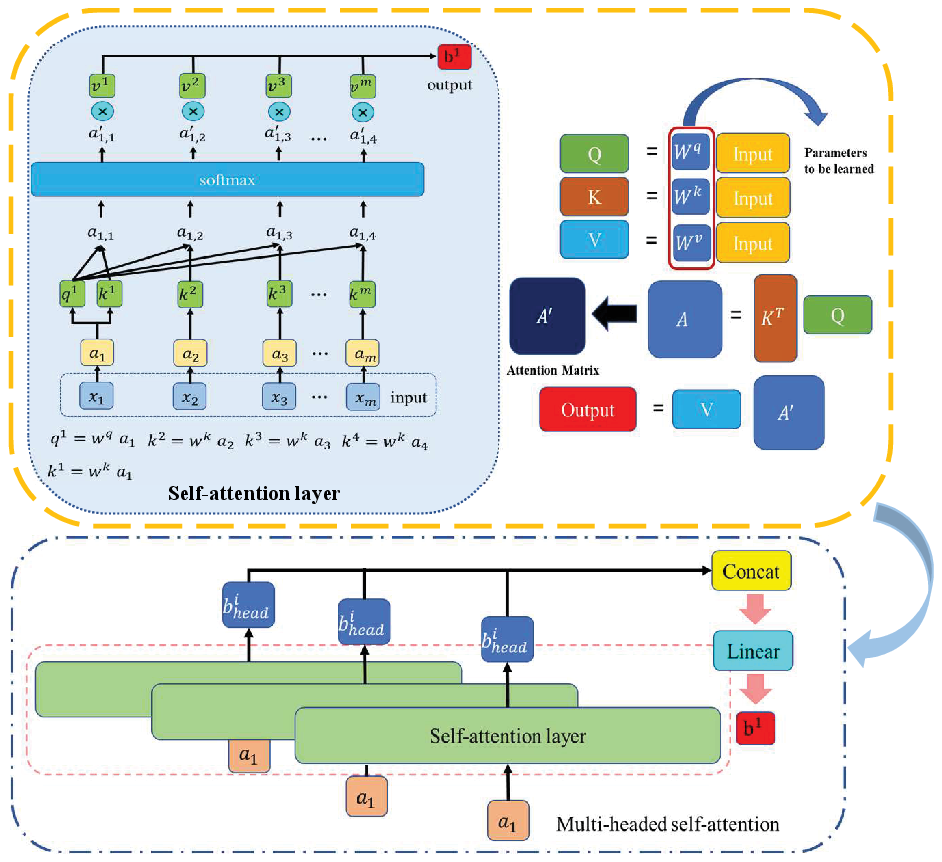}
	\caption{The principle of the multi-head self-attention mechanism.}
	\label{fig4}
\end{figure*}

The self-attention mechanism constitutes a distinct variant of the attention mechanism, primarily employed for processing temporal data such as text, images, or time-series data. It enables the model to analyze the sequence in a manner that takes into account the inter-element relationships within the sequence. The self-attention mechanism calculates weights by evaluating the similarities between different elements in a sequence to determine how much an element is related to other elements. Specifically, a kind of similarity score is defined to describe the similarity between each element in the sequence and other elements. Then, based on these similarity scores, the weight of each element can be calculated, and a weighted representation can be further obtained by multiplying the weight with the corresponding element. In order to better capture the associations between different types of SSP, multiple attention heads are used in this paper as shown in figure~\ref{fig4}. 

The diagram presented in Figure \ref{fig4} illustrates the intricate workings of the multi-head self-attention mechanism, which is a fundamental building block in the architecture of transformer models, which  hinges on three integral elements: the query $\mathcal{Q}$, key $\mathcal{K}$, and value $\mathcal{V}$ vectors. These vectors are derived from the input data through linear transformations, as depicted in the upper right section of the diagram.

The query vector serves to assess the relevance or correlation with the key vectors. This correlation is quantified and utilized to compute a weighted sum, which amalgamates the contributions of the value vectors. The key vectors are instrumental in determining the attention distribution function, denoted as $a_i$ in the diagram, which signifies the importance or weight of each element within the input sequence. In contrast, the value vectors encapsulate the specific features or numeric values associated with each element, playing a crucial role in the aggregation of information.

The operation of the mechanism commences with the calculation of the dot - product between the query and every key vector. To avoid an excessive increase in the dot-products that could trigger numerical instability, particularly when the dimensionality of the keys $d_k$ is substantial, each key is standardized through division by $\sqrt{d_k}$. This standardization stage is depicted in the diagram as an element of the attention computation procedure. Afterward, the softmax function is employed on the standardized dot - products to yield a collection of attention weights. Subsequently, these weights are utilized to calculate a weighted sum of the value vectors, as expressed by the equation:

\begin{equation}
    \text{Att}(\mathcal{Q}, \mathcal{K}, \mathcal{V}) = \text{softmax}\left(\frac{\mathcal{Q}\mathcal{K}^\top}{\sqrt{d_k}}\right)\mathcal{V}
\end{equation}

This operation is carried out for every head within the attention mechanism, as illustrated in the lower part of the diagram. Each head focuses on distinct portions of the input sequence, enabling the model to grasp a wide variety of dependencies and subtleties in the data. Subsequently, the outputs of all heads are concatenated and undergo a linear transformation, culminating in the ultimate output of the attention mechanism.

Given an input vector $\mathbf{x} \in \mathbb{R}^{dim}$, represented as $\mathcal{X} = [\mathbf{x}_1, \mathbf{x}_2, \ldots, \mathbf{x}_M]$, where $dim$ signifies  the number of feature dimensions and $M$ indicates the number of sequence samples. Once a query vector  $\mathbf{q}$ is received, the self-attention function is formulated as:

\begin{equation}
\mathrm{att}(\mathcal{X}, \mathbf{q}) = \sum_{i=1}^M a_i \mathbf{x}_i = \sum_{i=1}^M \frac{\exp(\mathrm{s}(\mathbf{k}_i, \mathbf{q}))}{\sum_{j=1}^M \exp(\mathrm{s}(\mathbf{k}_j, \mathbf{q}))} \mathbf{v}_i,
\end{equation}
where $a_i$ is the $i$th attention weight, computed by the softmax function over the scaled dot product of the key vector $\mathbf{k}_i$ and the query vector $\mathbf{q}$. The function $\mathrm{s}(\mathbf{k}_i, \mathbf{q})$ represents the similarity score between the key vector $\mathbf{k}_i$ and the query vector $\mathbf{q}$, which is typically calculated as the dot product $\mathbf{k}_i^\top \mathbf{q}$ divided by the square root of the key vector's dimension $\sqrt{d_k}$. The value vector $\mathbf{v}_i$ corresponds to the $i$th sample in the input sequence.

In a standard self-attention module, the multi-head attention mechanism employs multiple query vectors $\mathcal{Q}$ to select multiple sets of information from the input data in parallel. Specifically, each head independently computes the attention weights between the query vectors $\mathcal{Q}$, key vectors $\mathcal{K}$, and value vectors $\mathcal{V}$ to extract different subsets of information. The resultant representation of the multi-head self-attention mechanism is derived through the concatenation of results from each attention head, followed by a linear projection operation. The mathematical expression is:

\begin{equation}
\begin{aligned}
\mathrm{MultiHead}(\mathcal{Q},\mathcal{K},\mathcal{V})& =\mathrm{Concat}(\mathrm{head}_1,\ldots,\mathrm{head}_\mathrm{h})\mathcal{W}^O, \\
\end{aligned}
\end{equation}
where $\mathrm{head_i} =\text{Attention}(\mathcal{QW}_i^Q,\mathcal{KW}_i^K,\mathcal{VW}_i^V)$, $\mathcal{W}_{i}^{Q}\in\mathbb{R}^{d_{k}},\mathcal{W}_{i}^{K}\in\mathbb{R}^{d_{k}},\mathcal{W}_{i}^{V}\in\mathbb{R}^{d_{v}}$, and $\mathcal{W}^o\in\mathbb{R}^{hd_v}$ is a projection of the parameter matrix.

\section{Results and discussions}
To test the effectiveness of the proposed SA-MDF-CNN model, the monthly average fusion data from July 2015 to 2019 were selected as input training data and the monthly average SSP from 2020 were set to be the output label data. Experimental results were compared with other widely used methods for constructing spatial sound velocity distribution, including  spatial interpolation (SITP) and mean values (MEAN).  Additionally, we have incorporated new comparison algorithms: the convolutional neural network (CNN) method used in 2024 and the self-organising map (SOM) method used in 2025 respectively. The model parameter settings are given in Table~\ref{table1}.

\begin{table}[!htbp]
	\caption{\textbf{Parameter settings of SA-MDF-CNN}}
	\centering
	\begin{tabular}{cc}
		\toprule
		Parameter & Value  \\
		\midrule
            GPU & RTX 3090    \\  
		input size & [1976,6,8] \\
            self-attention & [8,32] \\
		filter size & [2,2] \\
		number of Channels & 256 \\
		number of filters &  256 \\
		convolution stride size & [1,1] \\
		pool size & [2,2] \\
		pooling stride  & 2 \\
		full connected output size & 1976 \\
            minimum batch size & 16\\
            maximum epochs & 100 \\
            learning rate & $10^{-3}$\\
            drop factor of learning rate & 0.1 \\
            drop period of learning rate &20 \\
		\bottomrule
	\end{tabular}
	\label{table1}
\end{table}


\begin{figure}[!htbp]
	\centering
	\begin{minipage}[t]{0.23\textwidth} 
		\centering
		\includegraphics[width=\linewidth]{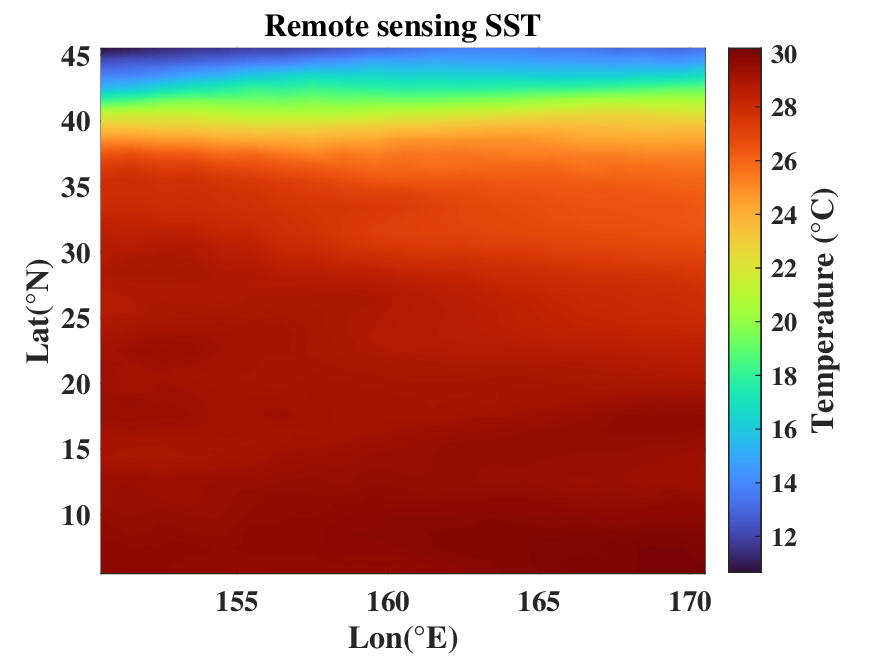} 
		 \caption{Remote sensing SST data.} 
		\label{fig5} 
	\end{minipage}
	\hfill 
	\begin{minipage}[t]{0.24\textwidth}
		\centering
		\includegraphics[width=\linewidth]{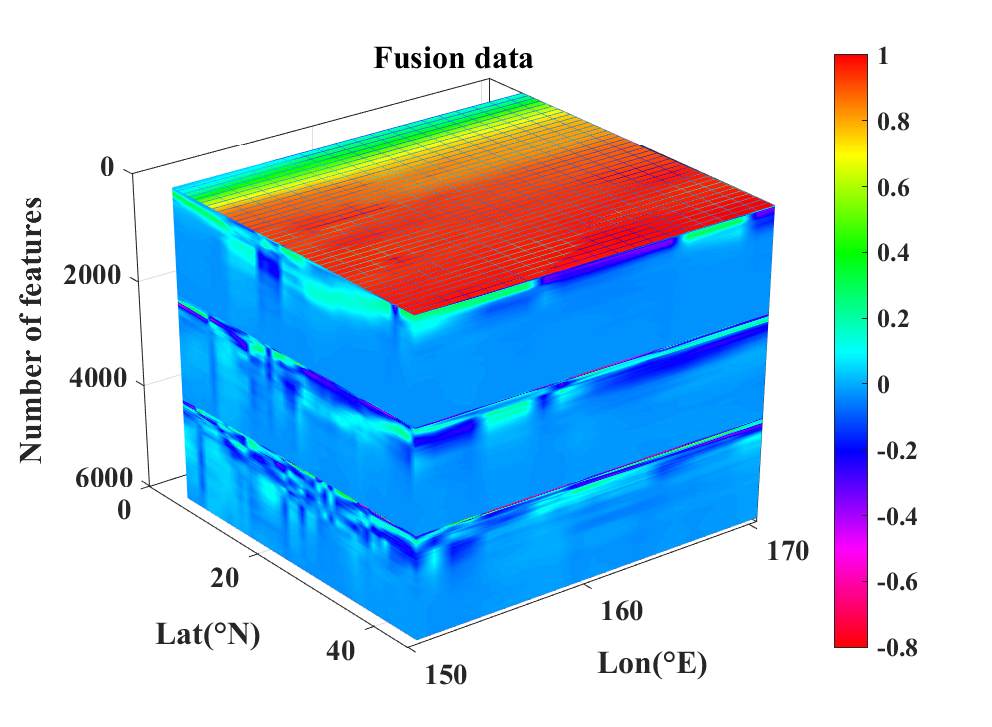}
		 \caption{Fusion data of remote sensing SST data, latitude and longitude coordinates, and EOF feature vectors.}
		\label{fig6}
	\end{minipage}
\end{figure}

To generate the fusion data necessary for evaluating the accuracy of the recommended model in estimating the sound velocity field, real-time remote sensing sea surface temperature (SST) data from multiple locations and depths were combined with the EOF feature vector to reconstruct the sound velocity field. As illustrated in Figure \ref{fig5}, the remote sensing SST data are presented. Moreover, Figure \ref{fig6} presents a visual depiction of the fused data.


\begin{figure*}[!htbp]
	\centering
	\includegraphics[width=\textwidth]{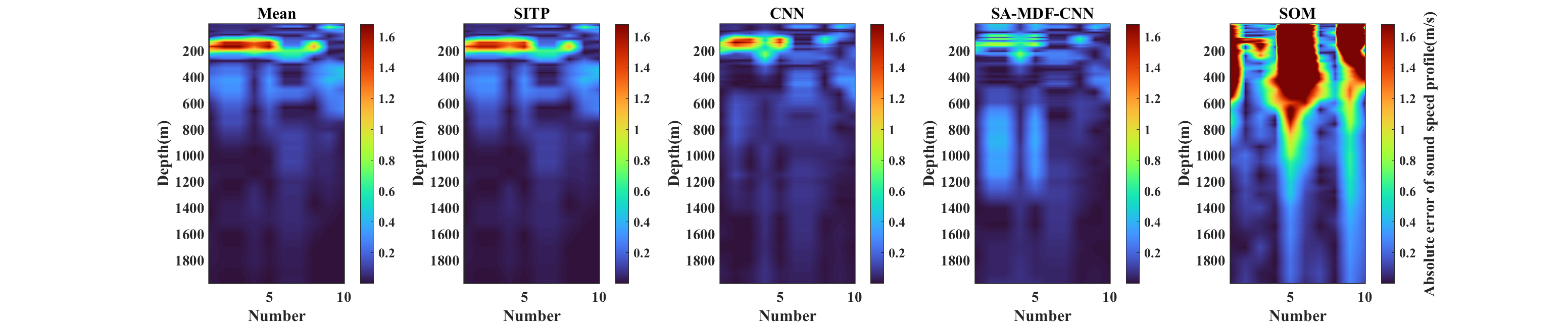}
	\caption{Centralised test Real-time SSP estimation error distribution for different algorithms at 1976 depth layers, where (a) is the mean value method, (b) is the spatial interpolation algorithm, (c) is the CNN algorithm,(d) is the SA-MDF-CNN algorithm, and (e)is the SOM algorithm}
	\label{fig7}
\end{figure*}
\begin{table}[htb]
	\caption{\textbf{Comparison of RMSE at the same depth:1975 meters.}}
	\centering
	\begin{tabular}{ccccccc}
		\toprule
		\multirow{2}{*}{\textbf{Number}} & \multirow{2}{*}{\textbf{Location}}& \multicolumn{5}{c}{\textbf{RMSE of different algorithms (m/s)}} \\ 
		\cmidrule(lr){3-7} 
		& (°N °E) &\textbf{Our} &\textbf{CNN} & \textbf{SITP} &  \textbf{Mean} &\textbf{SOM}  \\  
		\midrule
		1 & 7.5 156.5 & 0.1125 & 0.1317 & 0.2456 & 0.2591 & 0.4418 \\
		2 & 7.5 151.5 & 0.2599 & 0.3084 & 0.3182 & 0.3377 &0.4770\\
		3 & 6.5 157.5 & 0.1293 & 0.1402  & 0.1730 & 0.1812 & 0.6944\\
		4 & 8.5 163.5 & 0.1152 & 0.1219 &0.2775 & 0.2903&1.0581\\
		5 & 11.5 165.5 & 0.2090 &0.2149  & 0.2709 & 0.2842&0.6599\\
		6 & 24.5 162.5 & 0.0787 &0.1115 & 0.1232 & 0.1281&0.3518\\
		7 & 26.5 163.5 & 0.1181 & 0.1378 &0.1293 & 0.1367&0.3723 \\
		8 & 27.5 153.5 & 0.1039 & 0.1239 &0.1532 & 0.1621&0.7575\\
		9 & 7.5 160.5 & 0.1945 & 0.2050 & 0.4025 & 0.4229 &1.2434\\
		\midrule
		\textbf{Average}  & & \textbf{ 0.1468} & \textbf{0.1661} & \textbf{0.2326} & \textbf{0.2447}& \textbf{0.6729}\\
		\bottomrule
	\end{tabular}
	\label{table2}
\end{table}

\begin{figure}[!htbp]
	\centering
	\subfloat[]{\includegraphics[width = 0.24\textwidth]{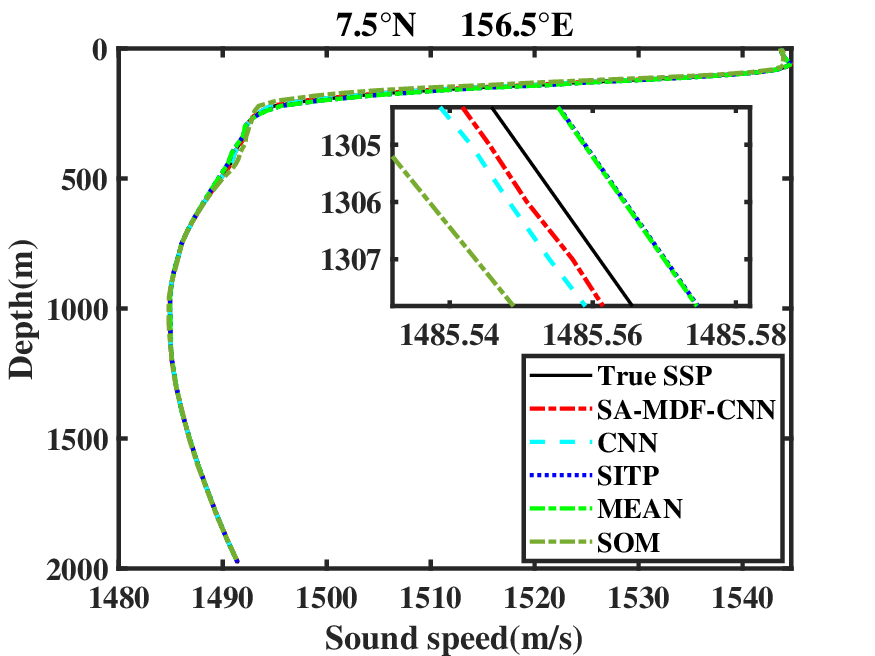}}
	\subfloat[]{\includegraphics[width = 0.24\textwidth]{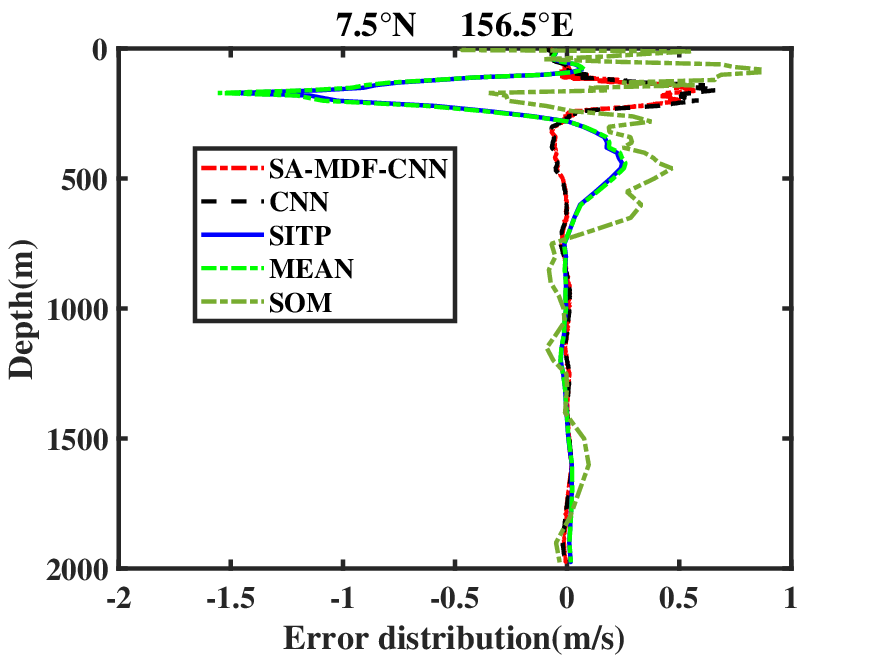}}
	\caption{A comparison example of real-time estimated SSP results of different algorithms with depths ranging from 0 to 1975 meters, where (a) is the comparison of estimated results and real SSP, and (b) is the comparison of error distribution between the estimated results of different algorithms and the real SSP.}
	\label{fig8}
\end{figure}

The distribution of absolute errors between real-time SSP and forecast results of SA-MDF-CNN algorithm with depths ranging from 0 to 1975 meters within $7^\circ N$ to $28^\circ N$, $150^\circ E$ to $165^\circ E$ are compared with those of CNN, SITP, SOM and mean values method as shown in figure~\ref{fig7}. From the intuitive representation of the sound velocity heat-map, the sound velocity distribution estimated by the SA-MDF-CNN and CNN models exhibits a substantial enhancement in accuracy relative to the SITP, SOM and Mean value methods, as evidenced by the results, and the absolute error between the real-time predicted SSP and the true SSP of the proposed model is minimised. A comprehensive evaluation of the accuracy performance of various algorithms across different locations at depths ranging from 0 to 1975 meters is presented in Table~\ref{table2}. Intuitive observation shows that the average RMSE results of SA-MDF-CNN, CNN, SITP, and Mean value methods are 0.1468, 0.1661, 0.2326, and 0.2447, respectively. Notably, the RMSE of SA-MDF-CNN for real-time SSP estimation is approximately 12\% lower than that of CNN and approximately 30-40\% lower than that of traditional SITP and mean value methods. Figure~\ref{fig8} gives a comparison of sound velocity estimation curves at $7.5^\circ N, 156.5^\circ E$. In figure~\ref{fig8} (a), the curve of SA-MDF-CNN is the closest to the real SSP, and in figure~\ref{fig8} (b), the error mainly manifests in the shallow ocean area within 500 meters because the sound velocity in shallow water is more severely and irregularly affected by temperature. However, among all methods the error disturbance of SA-MDF-CNN is the smallest.

\begin{figure*}[htbp]
	\centering
	\subfloat[]{\includegraphics[width = 0.30\textwidth]{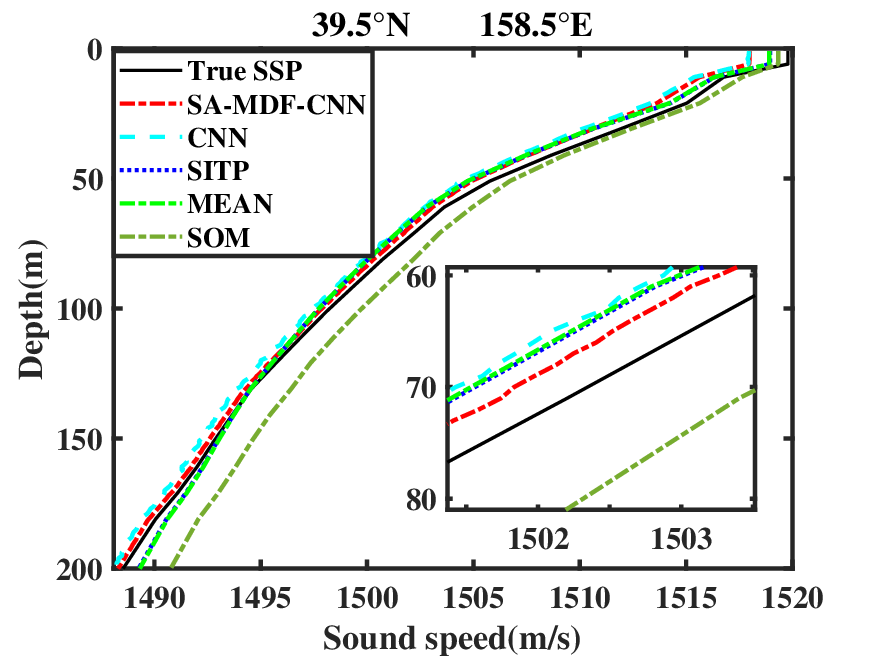}}\quad
	\subfloat[]{\includegraphics[width = 0.30\textwidth]{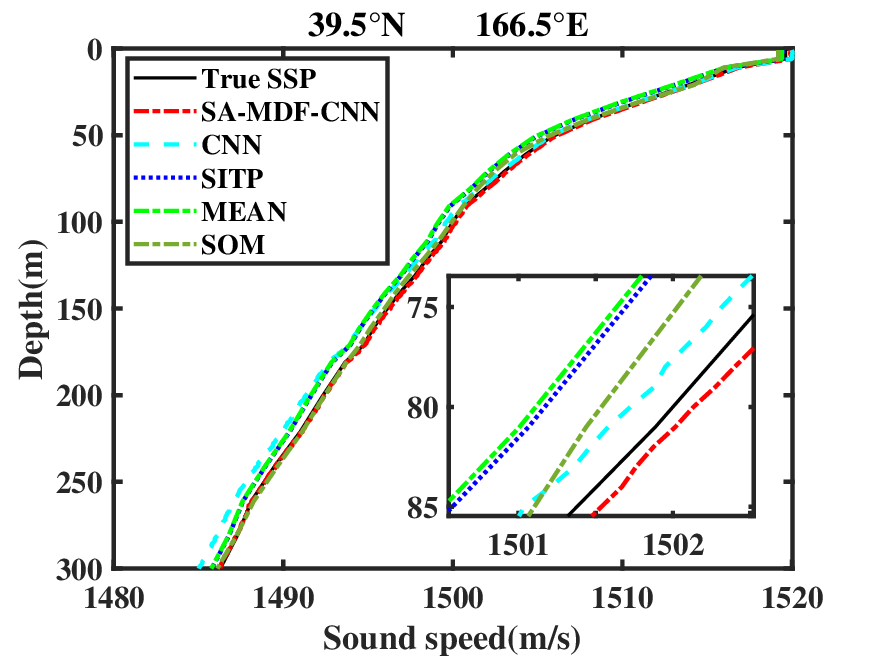}}\quad
	\subfloat[]{\includegraphics[width = 0.30\textwidth]{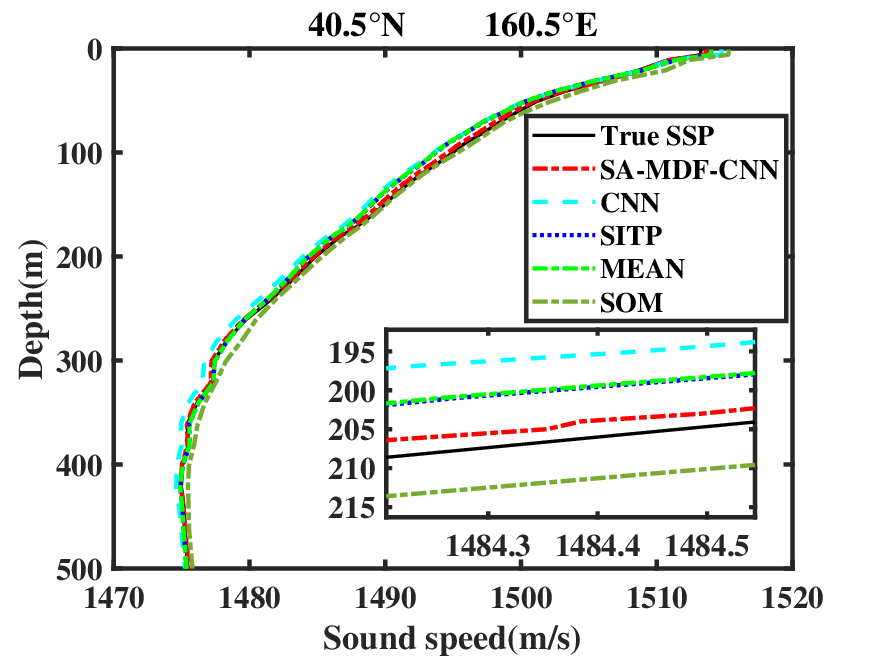}}\quad
	\subfloat[]{\includegraphics[width = 0.30\textwidth]{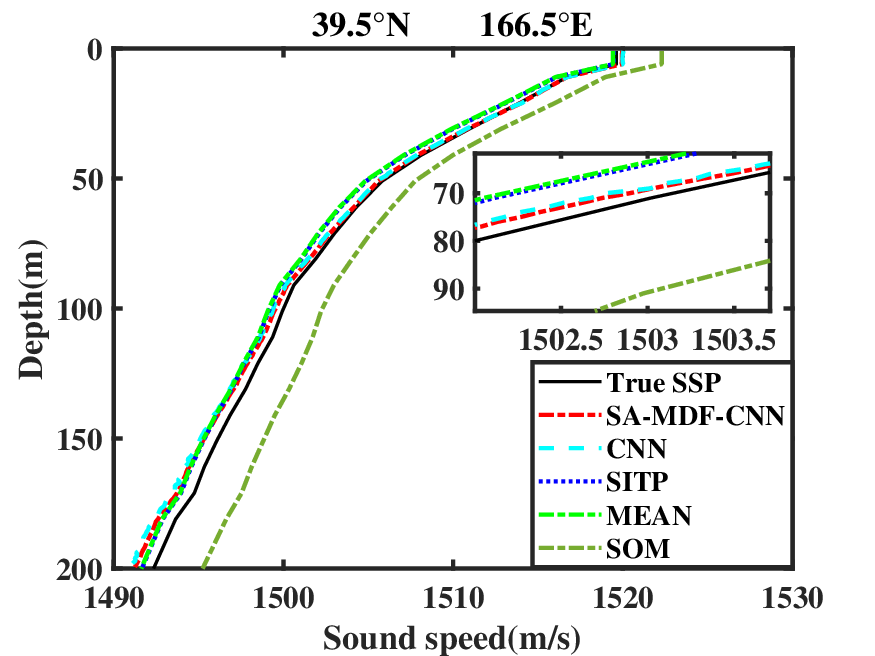}}\quad
	\subfloat[]{\includegraphics[width = 0.30\textwidth]{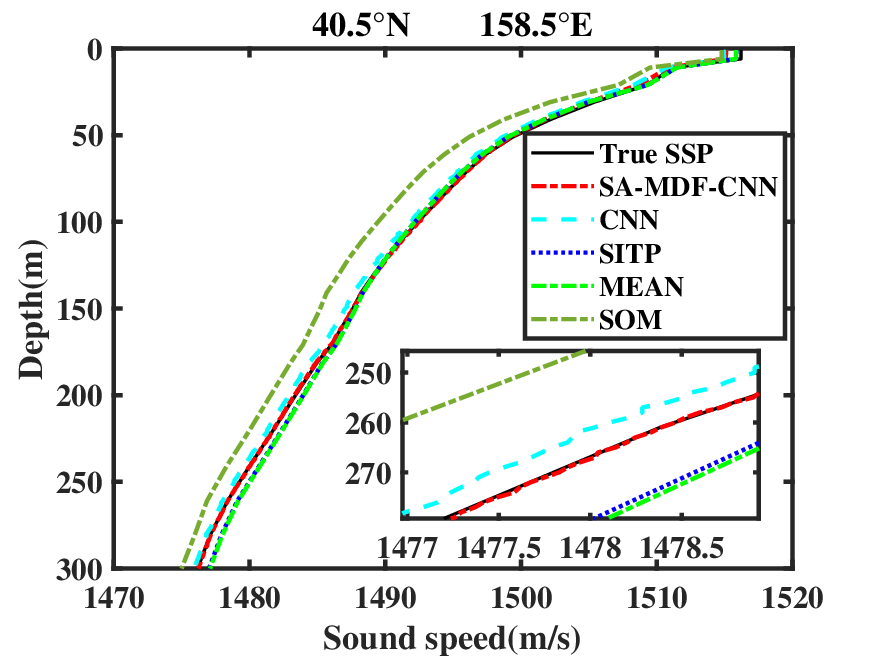}}\quad
	\subfloat[]{\includegraphics[width = 0.30\textwidth]{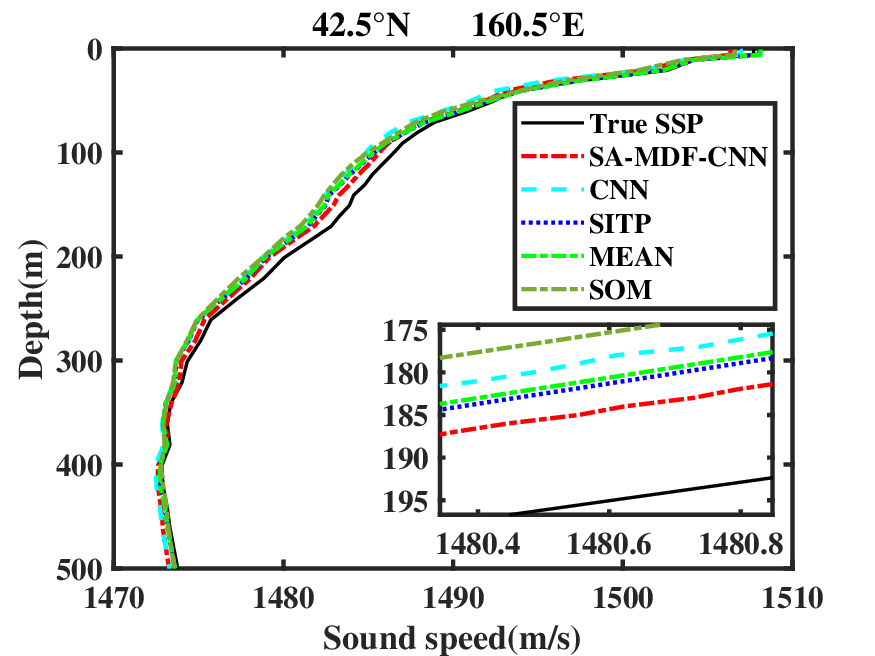}}\quad
	\caption{Comparison of estimation results of different algorithms at different shallow ocean depths.}
	\label{fig9}
\end{figure*}

\begin{table}[h]
	\centering
	\caption{Comparison of RMSE at different depth.}
	\begin{tabular}{cccccccc}
		\toprule
		\multirow{2}{*}{\textbf{Depth}} & \multirow{2}{*}{\textbf{Location}} & \multicolumn{5}{c}{\textbf{RMSE of different algorithms (m/s)}} \\
		\cmidrule(lr){3-7}
		& (°N °E) & \textbf{Our} & \textbf{CNN}  & \textbf{SITP}  & \textbf{Mean} & \textbf{SOM} \\
		\midrule
		\multirow{5}{*}{\textbf{200m}} & 6.5 159.5  & 0.6075& 0.6911 & 1.085 &1.1430&1.7247 \\
		& 7.5 156.5 & 0.2911& 0.3468 & 0.6688 &0.7035 &1.6666 \\
		& 7.5 161.5 & 0.3859 & 0.4092& 1.0190 & 1.0850&1.6325 \\
		& 9.5 164.5 & 0.1383  & 0.1775 & 0.3831 & 0.4018&1.8747 \\  
		& 13.5 169.5 &0.3115  & 0.3556 & 0.5565 & 0.5862& 1.1442\\  
		\multicolumn{2}{c}{\textbf{Average}} & \textbf{ 0.3469} & \textbf{0.3960} & \textbf{0.7425} &\textbf{0.7839}&\textbf{1.6085}\\
		\cmidrule(lr){3-7} \\
		\multirow{5}{*}{\textbf{300m}} & 11.5 165.5 & 0.4957 & 0.5155 & 0.6903 & 0.7240 & 2.7101\\
		& 12.5 152.5 & 0.1901 & 0.3242 & 0.3392 & 0.3542& 2.4136\\
		& 13.5 154.5 & 0.2889 & 0.3028 & 0.4363 & 0.4594& 1.2296\\
		& 14.5 158.5 & 0.1360 & 0.1411 & 0.3860 & 0.4088&0.6881 \\
		& 14.5 166.5 & 0.2943 & 0.3446 & 0.5869 & 0.6156 &1.3743\\
		\multicolumn{2}{c}{\textbf{Average}} & \textbf{0.2810} & \textbf{ 0.3256} & \textbf{0.4877} & \textbf{0.5124}& \textbf{1.6831}\\
		\cmidrule(lr){3-7} \\
		\multirow{5}{*}{\textbf{500m}} & 7.5 151.5 & 0.2293 & 0.2539 & 0.4839 & 0.5101  &0.9563\\
		& 6.5 157.5 & 0.2299 & 0.2613 & 0.3371 & 0.3530  &1.2761\\
		& 8.5 164.5 & 0.2943 & 0.3257 & 0.6033 & 0.6368  & 3.0348 \\
		& 11.5 156.5 & 0.1571 & 0.1625 & 0.3331 & 0.3481 &1.3084\\
		& 17.5 163.5 & 0.1133 & 0.1841 & 0.2544 & 0.2721& 1.7933\\
		\multicolumn{2}{c}{\textbf{Average}} & \textbf{0.2048} & \textbf{0.2375} & \textbf{0.4024} & \textbf{0.4240} & \textbf{1.6738}\\
		\bottomrule
		\label{table3}
	\end{tabular}
\end{table}

\begin{figure*}[!htbp]  
	\centering
	\subfloat[]{
		\includegraphics[width=\textwidth, keepaspectratio=false]{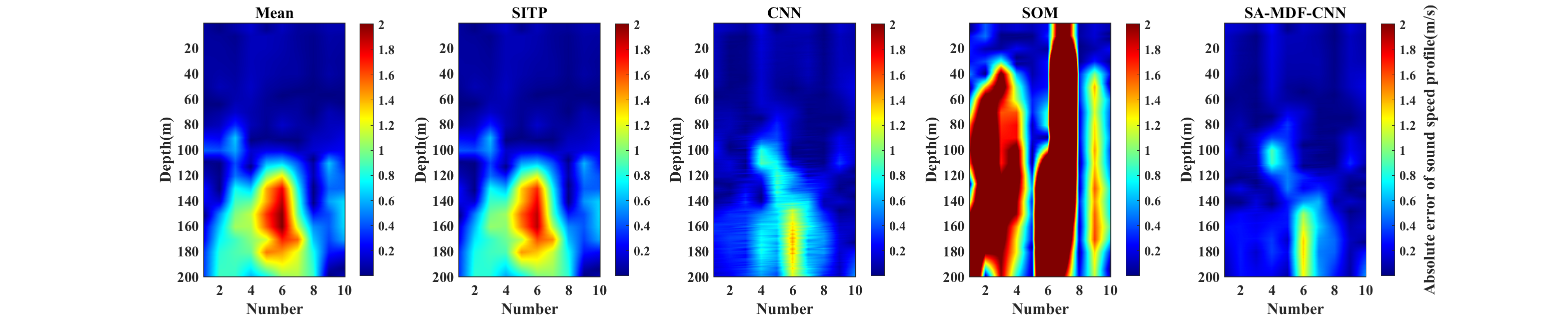}}\quad
	\subfloat[]{
		\includegraphics[width=\textwidth, keepaspectratio=false]{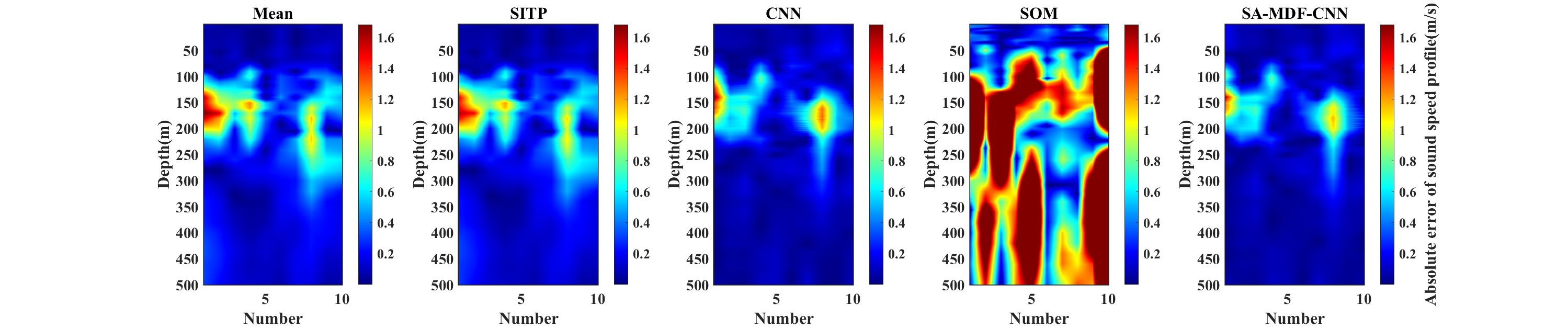}
	}
	\caption{Comparison of error distributions of real-time estimated SSPs for different algorithms at different shallow sea depths.}
	\label{fig10}
\end{figure*}

\begin{figure*}[htbp]
	\centering
	\subfloat[]{\includegraphics[width = 0.30\textwidth]{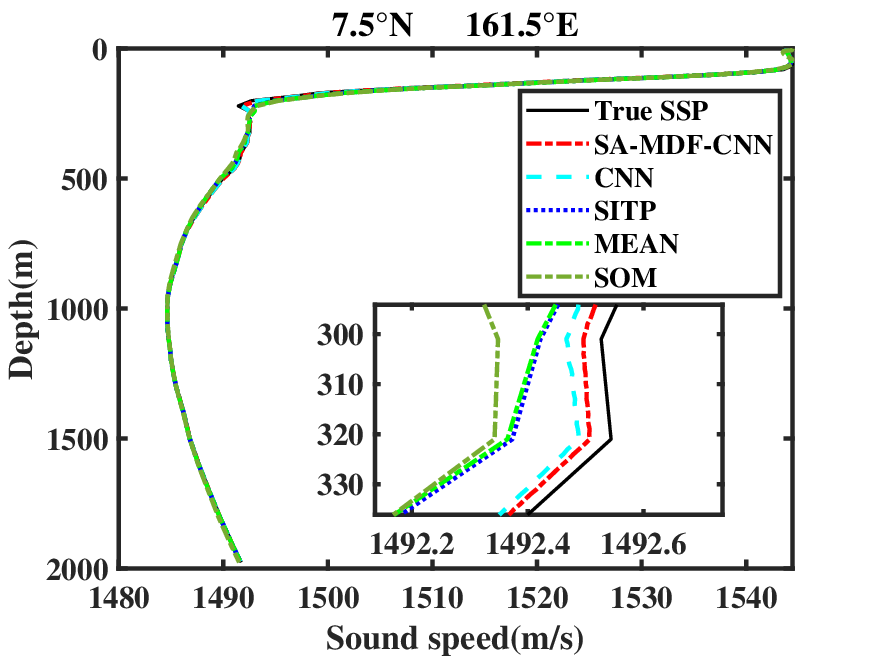}}\quad
	\subfloat[]{\includegraphics[width = 0.30\textwidth]{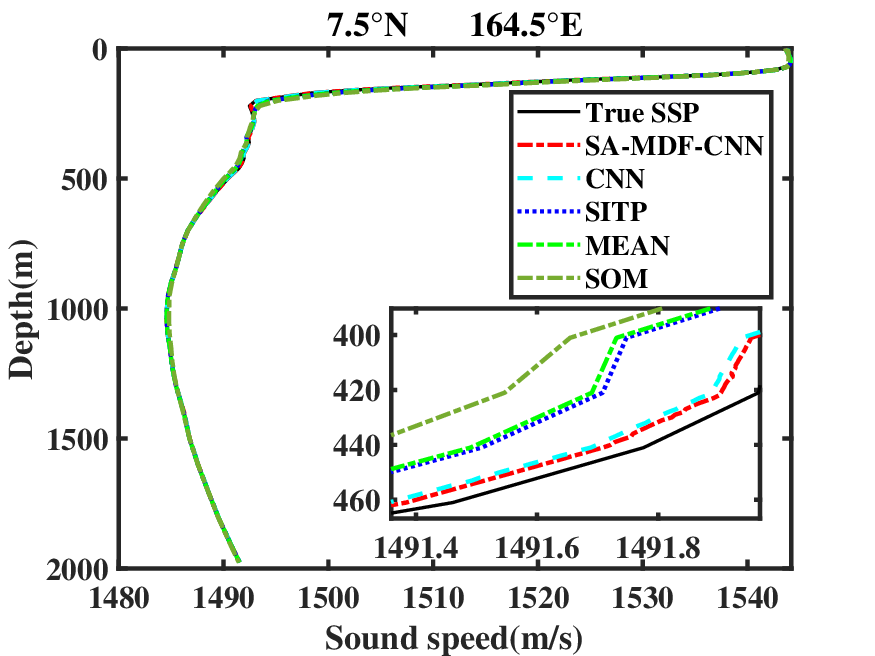}}\quad
	\subfloat[]{\includegraphics[width = 0.30\textwidth]{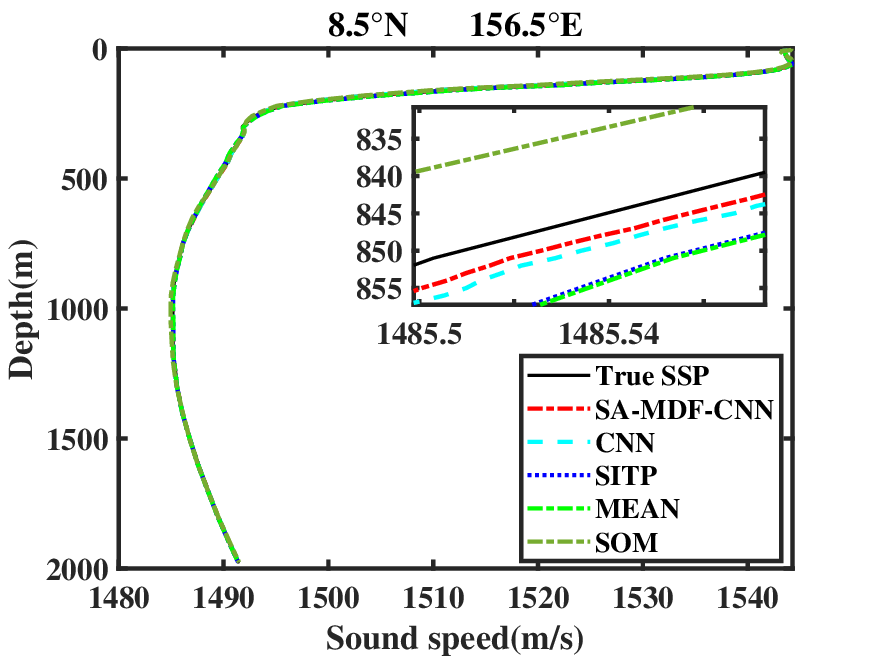}}\quad
	\subfloat[]{\includegraphics[width = 0.30\textwidth]{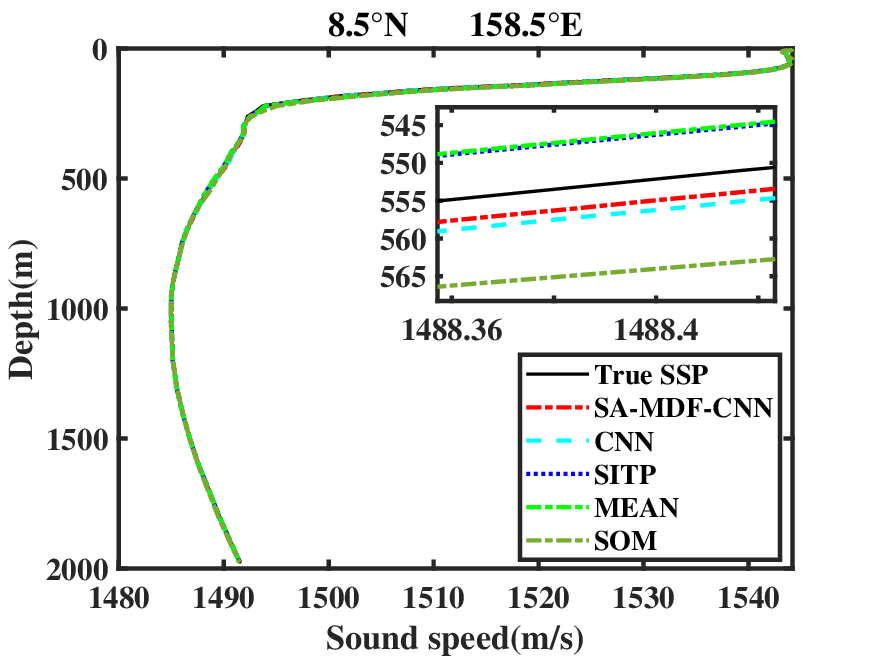}}\quad
	\subfloat[]{\includegraphics[width = 0.30\textwidth]{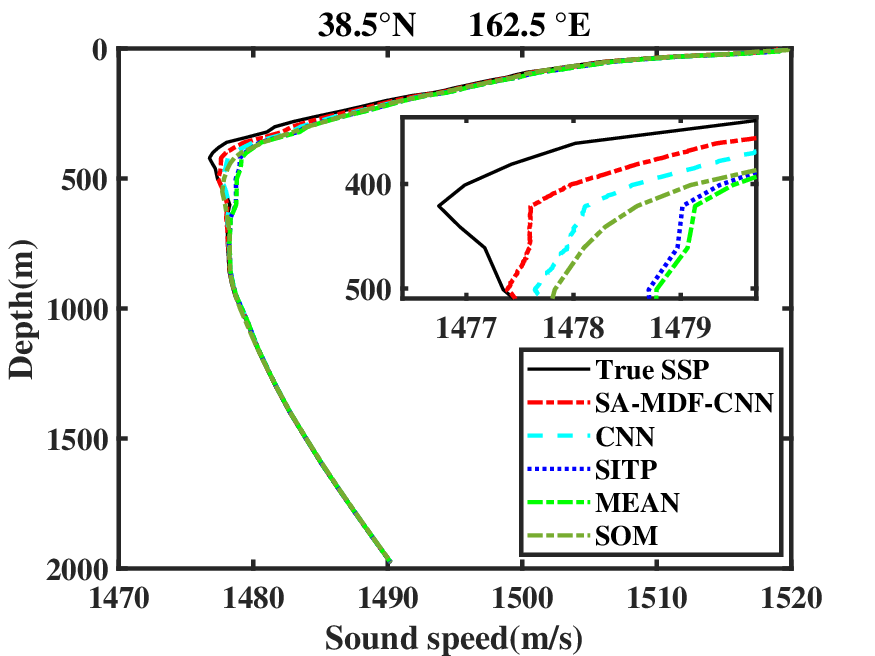}}\quad
	\subfloat[]{\includegraphics[width = 0.30\textwidth]{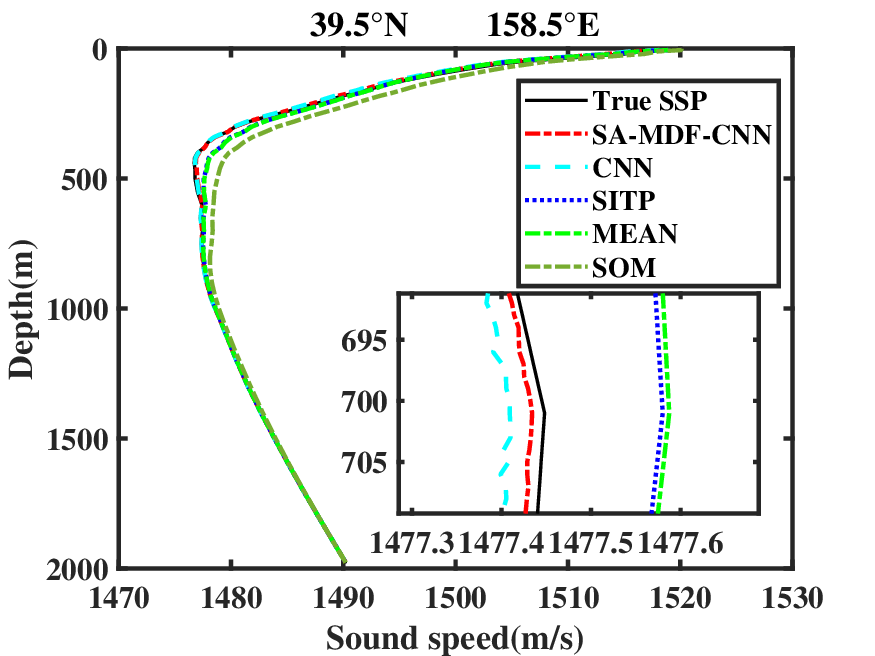}}\quad
	\caption{Comparison of different algorithms for unsmoothed SSP samples at different coordinates.}
	\label{fig11}
\end{figure*}

\begin{figure*}[!htbp]
	\centering
	\includegraphics[width=\textwidth]{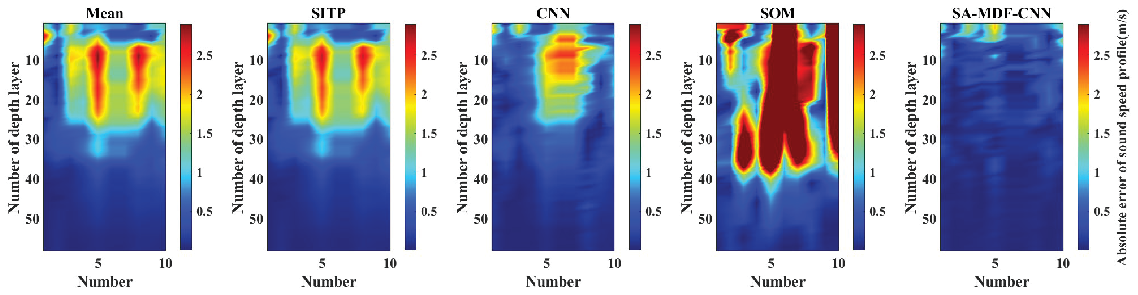}
	\caption{Concentrate on testing the absolute error distribution of non-smooth SSP under different algorithms at 58 depth layers.}
	\label{fig12}
\end{figure*}

\begin{figure*}[!htbp]
	\centering
	\subfloat[]{\includegraphics[width = 0.4\textwidth]{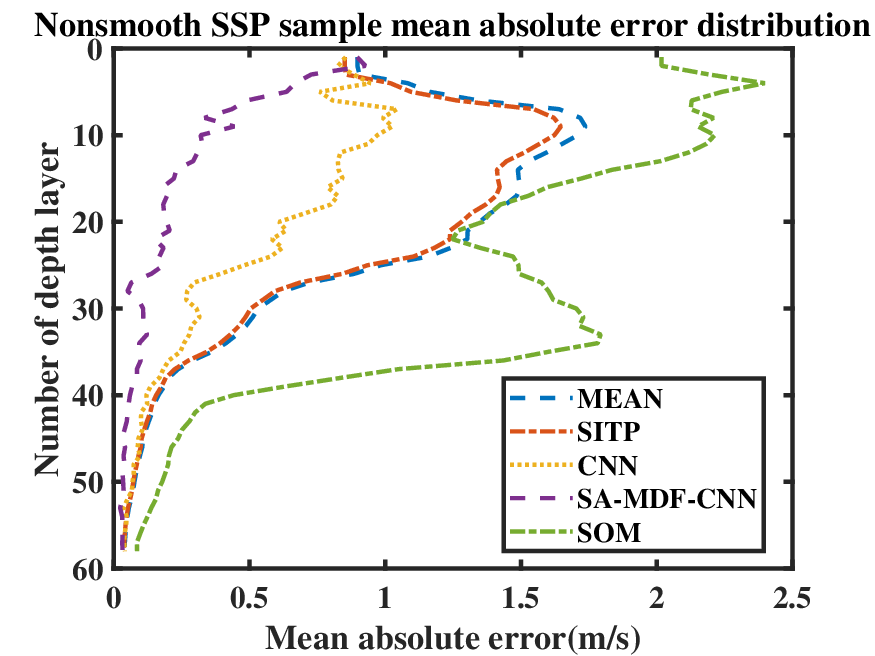}}\quad
	\subfloat[]{\includegraphics[width =0.5\textwidth]{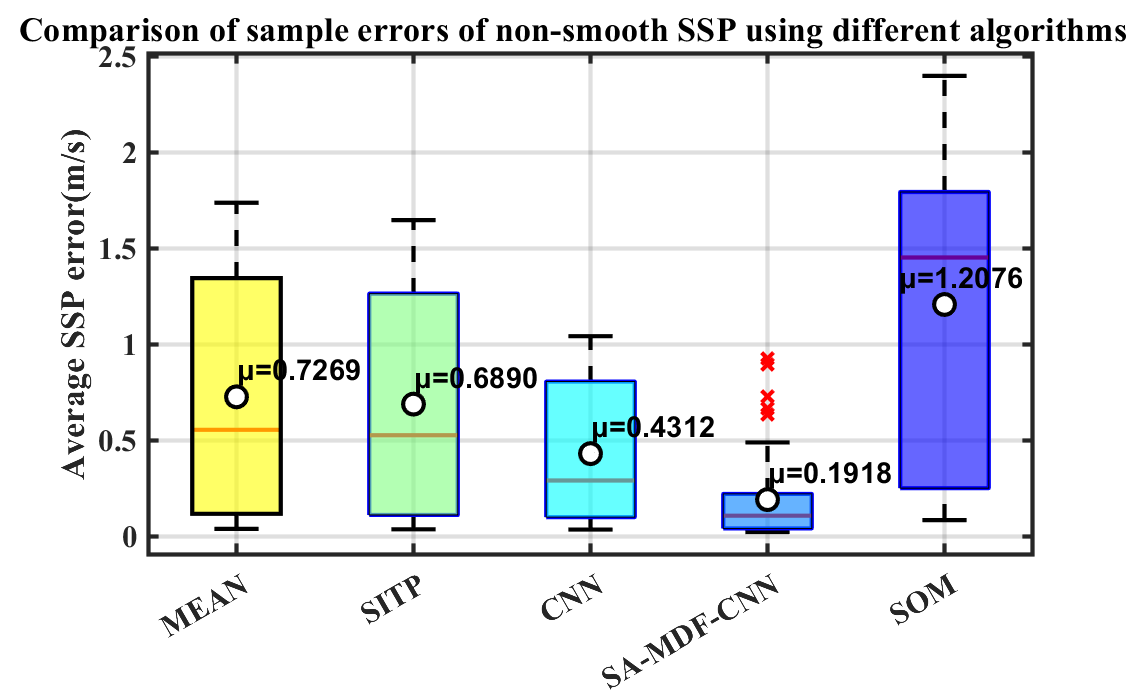}}
	\caption{Concentrate on testing the average absolute error of non-smooth SSP samples under different algorithms, where (a) shows the nonsmooth SSP sample mean absolute error distribution, and (b) shows the average absolute error box plot of non-smooth SSP under 58 depth layers under different algorithms.}
	\label{fig13}
\end{figure*}

\begin{figure*}[!htbp]
	\centering
	\includegraphics[width = \textwidth]{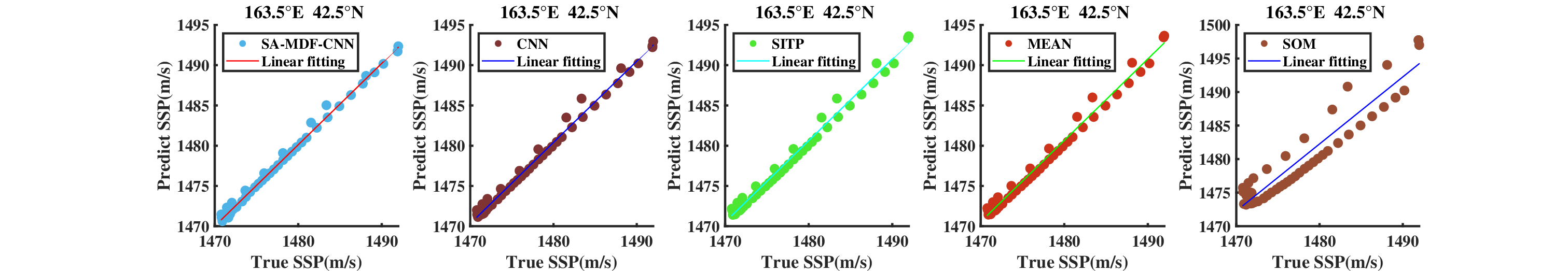}
	\caption{Comparison of scatter plots fitted by different algorithms at the same position for non-smooth SSP samples.}
	\label{fig14}
\end{figure*}

\begin{table*}[h]
	\centering
	\caption{The comparison of RMSE of different algorithms at different positions of non-smooth SSP samples under different grid conditions.}
	\begin{tabular}{ccccccccccccc}
		\toprule
		& \multicolumn{8}{c}{\textbf{RMSE of different algorithms (m/s)}} \\
		\cmidrule(lr){1-12}
\multirow{2}{*}{\textbf{Number}} & \multirow{2}{*}{\textbf{Location}} & \multicolumn{2}{c}{\textbf{SA-MDF-CNN}} & \multicolumn{2}{c}{\textbf{CNN}} & \multicolumn{2}{c}{\textbf{SITP}} & \multicolumn{2}{c}{\textbf{Mean method}} & \multicolumn{2}{c}{\textbf{SOM}} \\
	\cmidrule(lr){3-4} \cmidrule(lr){5-6} \cmidrule(lr){7-8} \cmidrule(lr){9-10} \cmidrule(lr){11-12}
		&& \textbf{4 grid} & \textbf{8 grid} & \textbf{4 grid} & \textbf{8 grid} & \textbf{4 grid} & \textbf{8 grid} & \textbf{4 grid} & \textbf{8 grid}  & \textbf{4 grid} & \textbf{8 grid}  \\
		\midrule
         1& 43.5°N 155.5°E  &0.2352 &0.4375&0.6577&0.5434&0.4531&0.6780 &0.4531&0.7251  &6.2802& 5.4812 \\
		 2 &43.5°N 158.5°E & 0.1116&0.1520&0.3344&0.2838&0.2084&0.2686&0.2084&0.2856   &1.8380& 2.3731\\
		 3	&43.5°N 165.5°E & 0.4191&0.3961&0.5127& 0.4797&0.8470&1.0420&0.8470&1.0820  &1.2565&2.4479\\
		 4	&43.5°N 166.5°E & 0.3664&0.2912&0.4969& 0.4826 &0.5680&0.8493&0.5680&0.9079 &0.6284 & 1.5560\\  
		 5	&43.5°N 167.5°E &0.5443 &0.5803 &1.0740&1.0800&1.1150&1.5130 &1.1150&1.5960 &0.2906&2.6898\\
		 6	&43.5°N 169.5°E & 0.2538 &0.1818 &1.2170&1.2200&0.5951&0.8829&0.5951&0.9426 & 1.2752 & 1.4193\\  
	  	 7 &44.5°N 155.5°E &0.2068 &0.1885&0.5322& 0.3578&0.6353&0.8935&0.6353&0.9475  &1.1789& 0.6226\\ 
		 8	&44.5°N 156.5°E & 0.2232 &0.2827&0.8552 &0.7471&1.0540&1.3390 &1.0540& 1.3980  &0.4634 & 1.2908\\ 
		 9	&44.5°N 157.5°E & 0.3117&0.2981&0.3962 &0.3833&0.6105&0.8854&0.6105&0.9425 &  7.3642& 1.3394\\
		 10	&42.5°N 166.5°E &0.3140 &0.4509& 0.5259&0.5100&0.8270&1.0570&0.8270& 1.1050  & 0.1710& 0.8194\\ 
		 11	&41.5°N 162.5°E &0.7789 &0.6077& 1.5290&1.4860&0.9531&1.2190&0.9531& 1.2740  & 1.6708 &1.8373\\ 
		 12	&42.5°N 162.5°E &0.6054 &0.4787& 0.7789&0.6055&0.5824&0.8591&0.5824& 0.9175  &0.9879 &1.0658\\ 
		 13	&42.5°N 163.5°E &0.4729 &0.4257& 0.7548&0.6508&0.6532&0.8562&0.6532& 0.8986 & 4.4901&3.1932\\
		\multicolumn{1}{c}{\textbf{Average}} && \textbf{0.3726}& \textbf{0.3670} & \textbf{0.7435}	& \textbf{0.6792} & \textbf{0.7002}& \textbf{0.9495} &\textbf{0.7002} &\textbf{1.0017} & \textbf{2.1458} &\textbf{2.0104}\\
		\bottomrule
		\label{table4}
	\end{tabular}
\end{table*}

To rigorously evaluate the precision and performance of the model for reconstructing acoustic velocity profiles in coastal marine environments, a comparative analysis was conducted between the instantaneous SSP reconstruction outcomes generated by the SA-MDF-CNN framework at 200 m, 300 m, and 500 m bathymetric layers and those produced by conventional CNN architectures, the SITP technique, SOM and statistical averaging approaches, as illustrated in Figure~\ref{fig9}. The magnified sections of individual subplots reveal that SA-MDF-CNN-derived SSP estimations at 200 m, 300 m, and 500 m depth layers demonstrate enhanced alignment with ground-truth SSP predictions when compared to conventional CNN architectures, SITP algorithms, SOM and mean method within identical marine regions. Specifically, an in-depth investigation into the efficacy of various algorithms within shallow ocean areas at varying depths and locations was conducted, with the real-time SSP estimation RMSE results compared and displayed in Table~\ref{table3}. The largest RMSE is observed at a depth of 200 meters, which can be attributed to the increased interference from real-time sea conditions in the shallow sea environment, such as wind speed, wave height, tsunami, and other factors. Concurrently, the experimental results indicate that the estimation accuracy of the proposed SA-MDF-CNN model at 200 meters depth is enhanced by approximately 13\% compared to CNN and by approximately 53\% compared to SITP and mean value method. At an ocean depth of 300 meters, the proposed algorithm demonstrates an improvement of about 14\% over CNN and about 42\% over SITP and mean value method. At 500 meters depth, the proposed model shows an improvement of about 14\% compared to CNN and about 49\% compared to SITP and mean value method. These experimental results collectively illustrate that the proposed model maintains significant precision advantages even in shallow sea environments of varying depths, thereby underscoring its applicability and effectiveness. Experimental findings reveal the SA-MDF-CNN architecture maintains operational efficacy across heterogeneous depth profiles within coastal marine conditions, while exhibiting RMSE relative to conventional methodologies, which means that the model still has good feature capture ability in shallow ocean scenes with significant changes in sound velocity. The average absolute error distribution comparison of real-time estimated SSP of different algorithms tested in the set of 200m and 500m depths is given in Figure \ref{fig10}. The error fluctuation of SA-MDF-CNN is the most stable, which indicates that SA-MDF-CNN still has better robust performance under shallow sea conditions with different depths.

To further evaluate the model performance, we performed statistical analysis of the reconstructed predictions using unsmoothed SSP samples from eastern Japan, where the dynamic variability is more pronounced. Figure \ref{fig11} displays the comparative results of different algorithms applied to unsmoothed SSP samples under varying coordinates. It is evident that even with unsmoothed SSP samples, the model maintains robust performance. Figure \ref{fig12} illustrates the absolute error distribution of real-time SSP predictions generated by different algorithms on a concentrated test set of unsmoothed 58-layer SSP samples. Table \ref{table4} compares the Root Mean Squared Error (RMSE) of different algorithms at various locations of unsmoothed SSP samples under distinct grid configurations. Experimental results indicate that 4-grid and 8-grid configurations exhibit comparable impacts on model performance, yet their applicability diverges. The 4-grid configuration is suitable for scenarios with minor SSP fluctuations in the target sea area. The 8-grid configuration demonstrates superior adaptation to significant SSP fluctuations in the target sea area, attributable to its enhanced capability in learning historical spatial distribution characteristics of SSP under such conditions. Figure \ref{fig13} presents the concentration test results of mean absolute errors (MAE) for unsmoothed SSP samples across different algorithms. Specifically: (a) Displays the MAE distribution of unsmoothed 58-layer SSP samples under various algorithms. (b) Illustrates the box plots of MAE for unsmoothed SSP across 58 depth layers utilizing different algorithms. The proposed SA-MDF-CNN algorithm demonstrates significantly superior performance in unsmoothed SSP prediction compared to conventional methods, achieving an MAE of 0.1811 m/s while maintaining stability throughout the full 58-layer profile. This substantiates the algorithm's enhanced robustness against oceanic environmental noise and SSP abrupt variations.

\begin{figure*}[htbp]
    \centering
    \subfloat[]{\includegraphics[width = 0.24\textwidth]{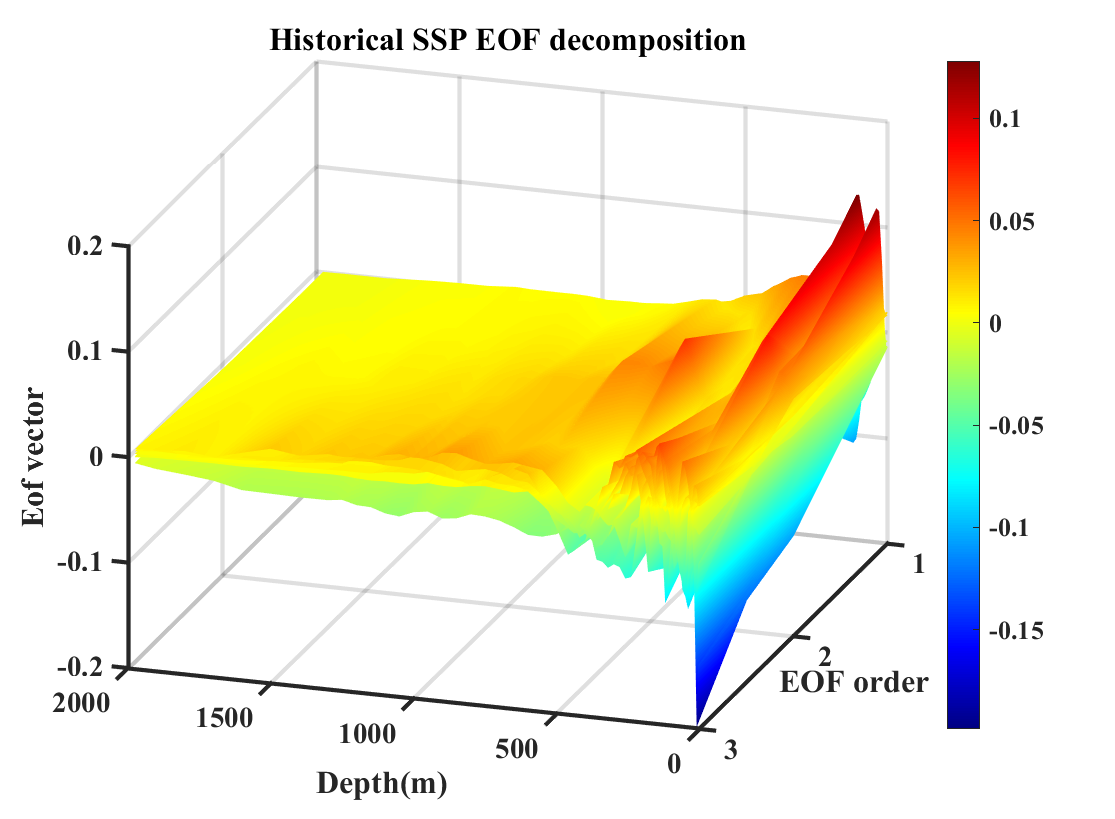}}
    \subfloat[]{\includegraphics[width = 0.24\textwidth]{fig15a.eps}}
    \subfloat[]{\includegraphics[width = 0.24\textwidth]{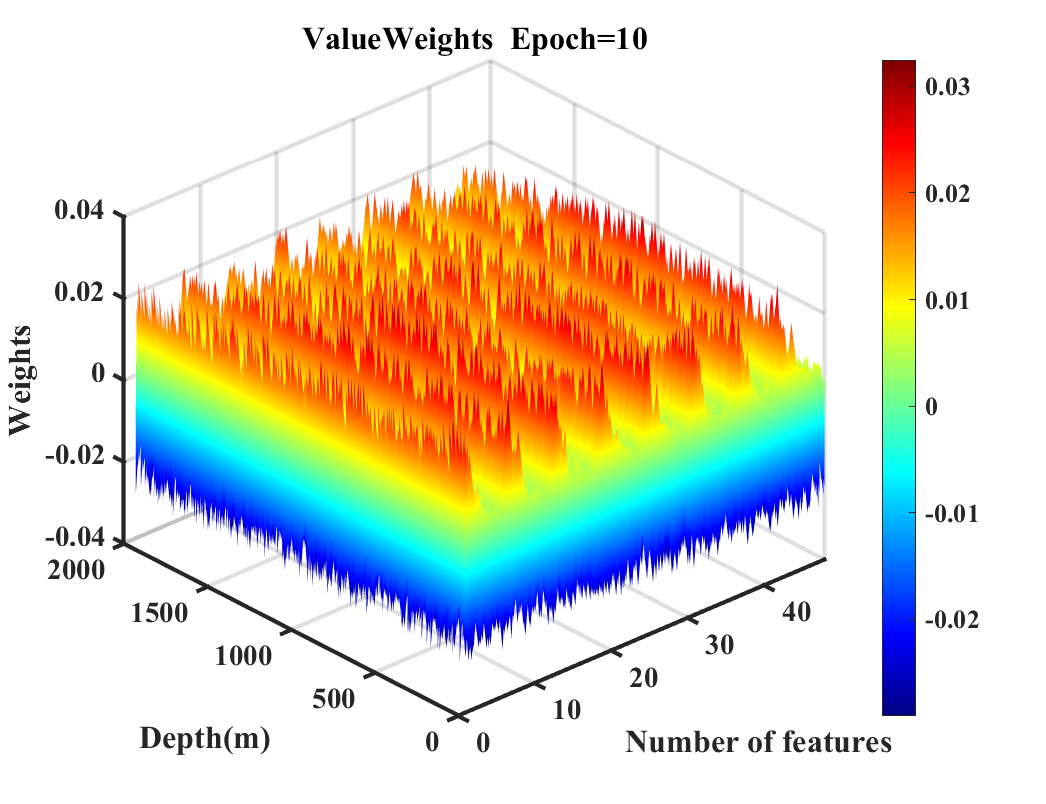}}
    \subfloat[]{\includegraphics[width = 0.24\textwidth]{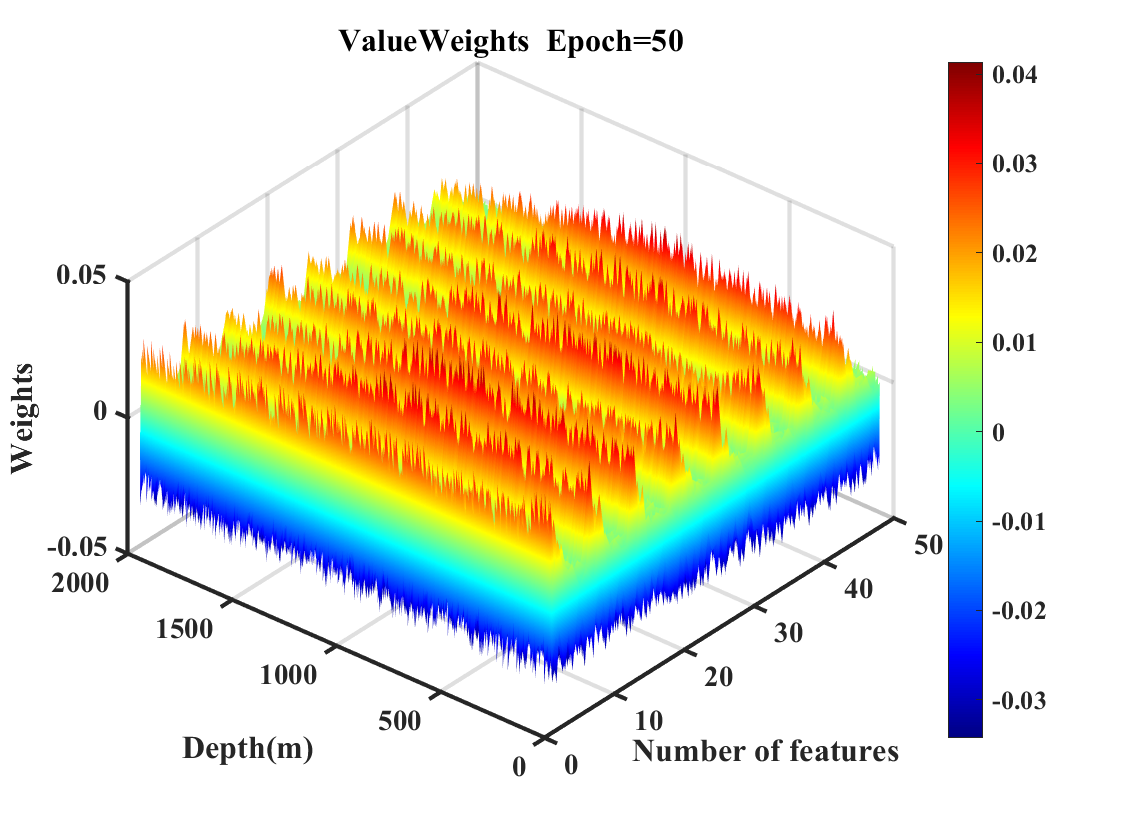}}
    \caption{Interpretability analysis of the SA-MDF-CNN model, where (a) is the first three order EOF the historical SSP at different positions (b) is the attention weight with Epoch=10, (c) is the attention weight with Epoch=50, and (d) is the attention weight at Epoch=10.}
    \label{fig15}
\end{figure*}

\subsection{Interpretability analysis}
To investigate why the proposed SA-MDF-CNN model has better performance compared to CNN, we conducted an experiment and visually represented the attention weights of the model as shown in figure~\ref{fig15}. Figure \ref{fig15} (a)  illustrates the primary three-order EOF eigenvectors corresponding to historical SSPs at multiple geospatial coordinates. Figure~\ref{fig15} (b), (c), and (d) correspond to the attention weights when the training epochs are 10, 50, and 100, respectively. Through observing the evolution trend of attention focus with increasing training times, we found that the SA-MDF-CNN model increasingly focuses on the acoustic velocity profile in shallow marine environments within 1000 meters. This trend aligns with the first three orders of EOF feature distribution shown in panel (a), although the EOF itself emphasizes the SSP distribution between 0-500 meters. The attention depth of the proposed model spans 0-1500 meters, which is attributed to the variation of SSP distribution under different latitude and longitude coordinates. To improve the capability of capturing global SSP distribution patterns, a multi-head self-attention mechanism was incorporated following the fusion of input data. This architectural enhancement demonstrates the robustness and validity of the introduced framework. Experiments based on the aforementioned public dataset demonstrate that the proposed SA-MDF-CNN model outperforms other algorithms in estimating sound velocity in both shallow and deep waters.

\subsection{Efficiency and stability}
\begin{table}[!htbp]
	\caption{\textbf{Comparison of Neural Networks}}
	\centering
	\begin{tabular}{cccc}
		\toprule
		Network & Number of Parameter & Train time & Test time \\
		\midrule
		CNN & 61.9M  & 40s    & 0.37s\\  
		SA-MDF-CNN & 149M  & 72s & 0.82s   \\
		\bottomrule
	\end{tabular}
	\label{tab4}
\end{table}

Although the accuracy performance in SSP estimation of SA-MDF-CNN is better than traditional methods, the disadvantage is that SA-MDF-CNN increases the number of parameters and computational overhead. Table~\ref{tab4} presents the computational complexity analysis and temporal efficiency comparison between our framework and conventional CNN architectures. The experimental results indicate that the introduced methodology requires increased parameter quantities and extended training and testing durations compared to CNN implementations, representing an essential computational investment for achieving enhanced predictive capabilities.

\begin{figure}[!htbp]
	\centering
	\includegraphics[width=0.4\textwidth]{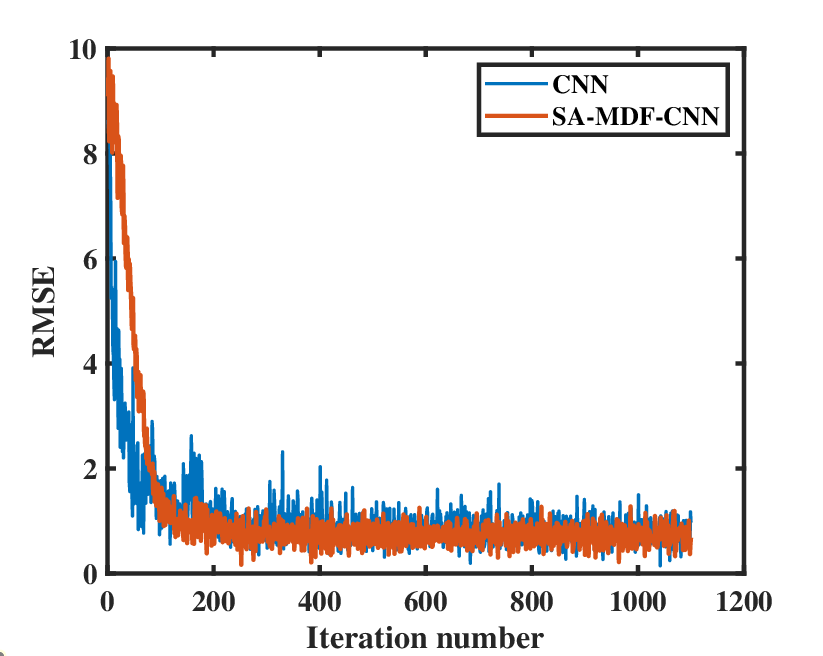}
	\caption{Comparison of RMSE convergence of different algorithms.}
	\label{fig16}
\end{figure}

To test the stability of the model, the convergence of SA-MDF-CNN and CNN with different numbers of iteration is given in figure~\ref{fig16}. Both SA-MDF-CNN and CNN can converge in less than 200 training iterations. The convergence of SA-MDF-CNN is smoother than that of CNN, which may be due to the multi-head self-attention mechanism focusing more on global features and therefore not causing significant parameter updates due to drastic changes in a small range of the data.

\begin{figure}[!htbp]
	\centering
	\subfloat[]{\includegraphics[width = 0.24\textwidth]{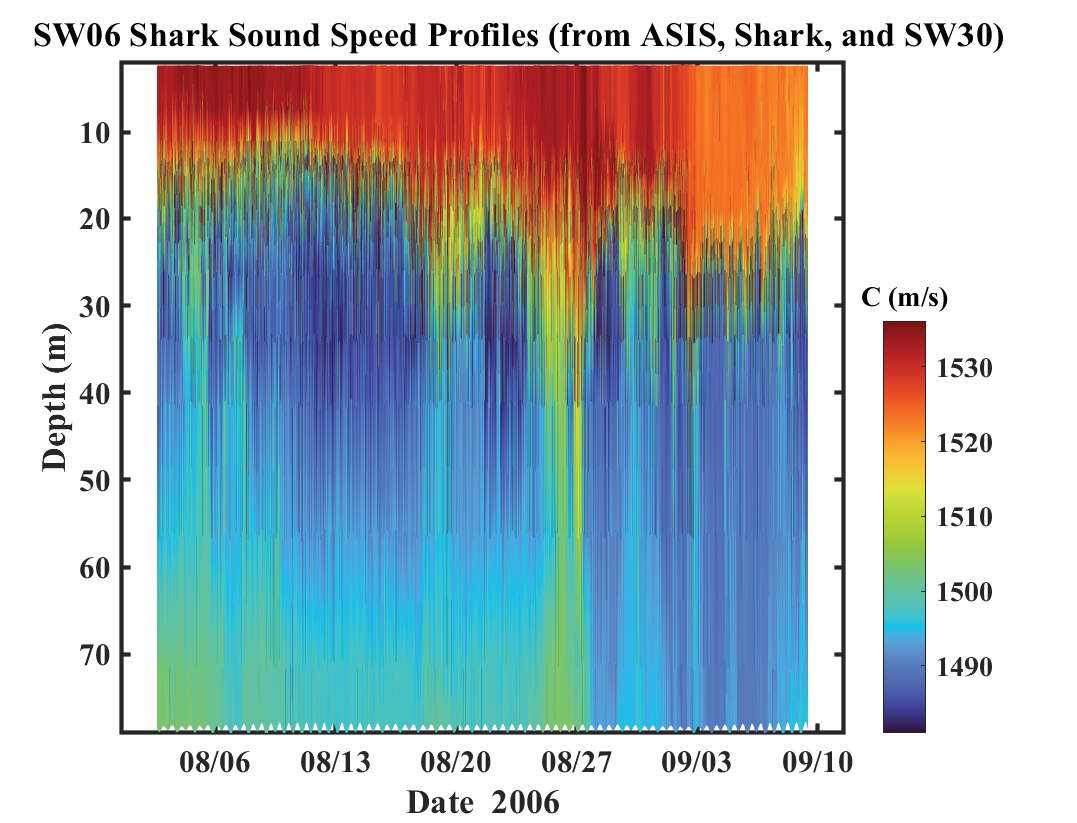}}
	\subfloat[]{\includegraphics[width = 0.24\textwidth]{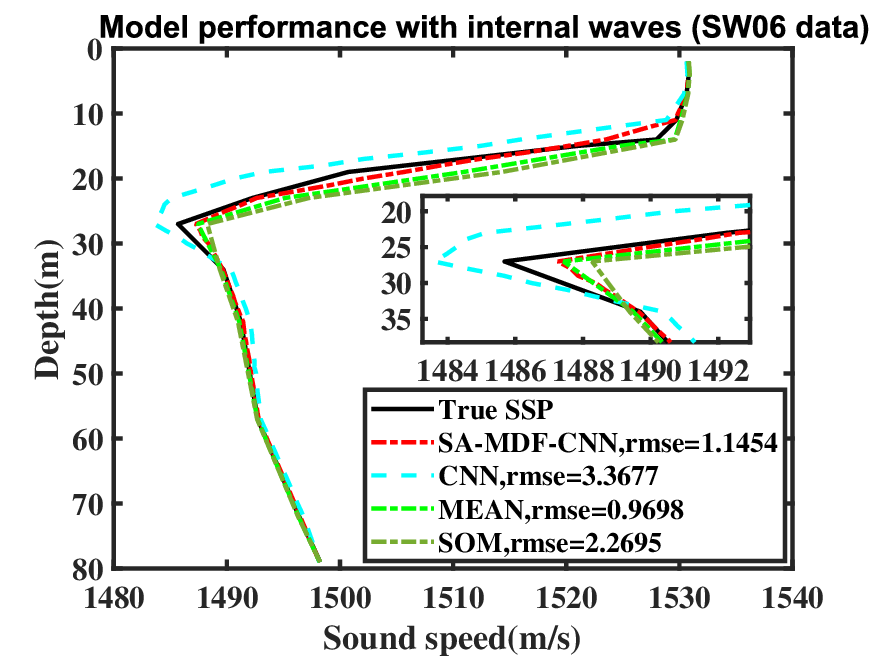}}
	\caption{Based on the test model results of SW06 SSP data, (a) is SW06 Shark SSP and (b) is the test results of different algorithms based on SW06 data.}
	\label{fig17}
\end{figure}

Figure \ref{fig17} presents validation results utilizing the public SW06 internal wave dataset to evaluate the model's capability in capturing mesoscale dynamics. Specifically: (a) Experimental SSP data with internal wave perturbations; (b) Real-time SSP predictions generated by different algorithms. The experimental results demonstrate that while the proposed algorithm exhibits certain limitations in capturing mesoscale dynamics within time series data, it significantly outperforms traditional CNN and SOM methods in prediction accuracy.

\section{Ocean experiments}

\begin{figure}[!htbp]
    \centering
    \subfloat[]{\includegraphics[width = 0.24\textwidth]{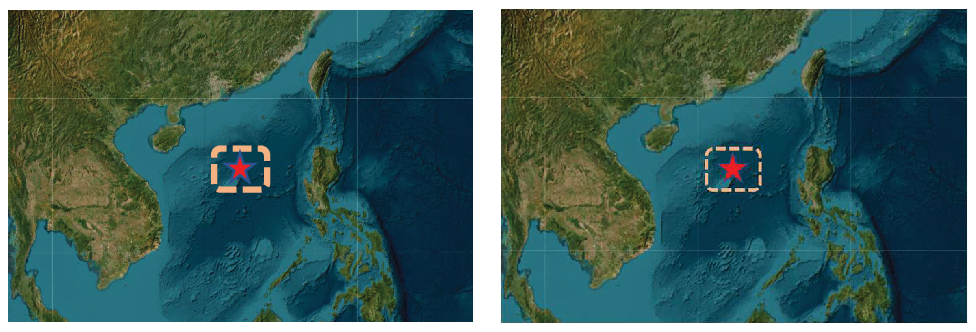}}
    \subfloat[]{\includegraphics[width = 0.24\textwidth]{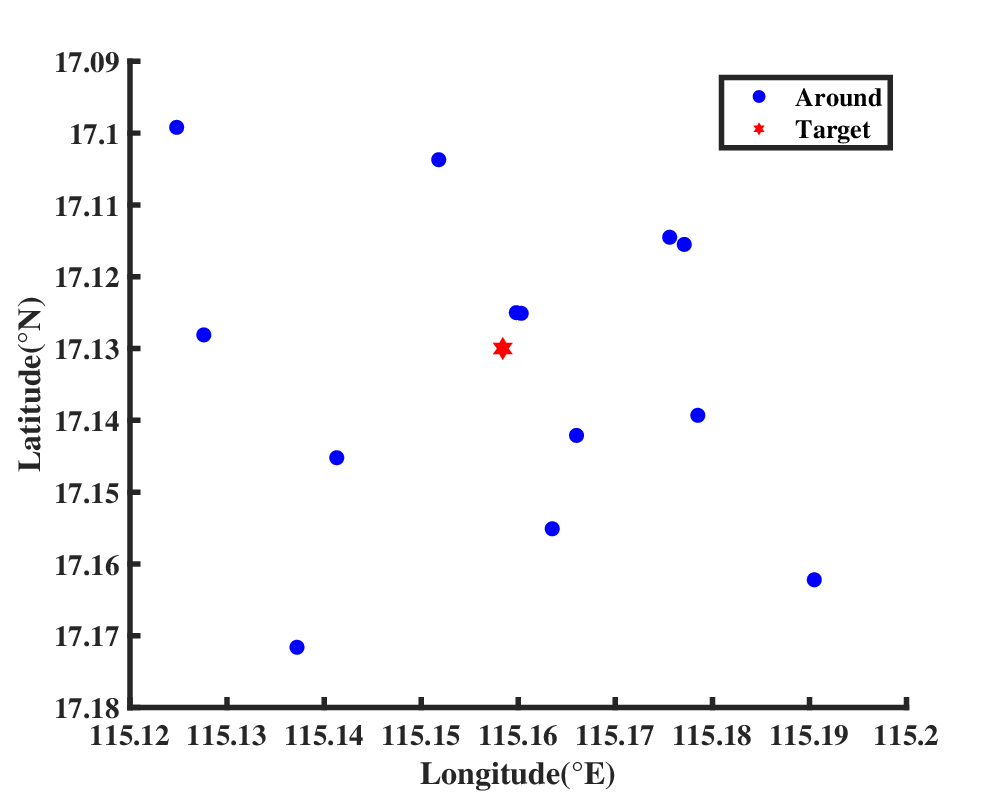}}
    \caption{Data sampling locations of ocean experiments.}
    \label{fig18}
\end{figure}

\begin{figure}[!htbp]
	\centering
	\subfloat[]{\includegraphics[width = 0.24\textwidth]{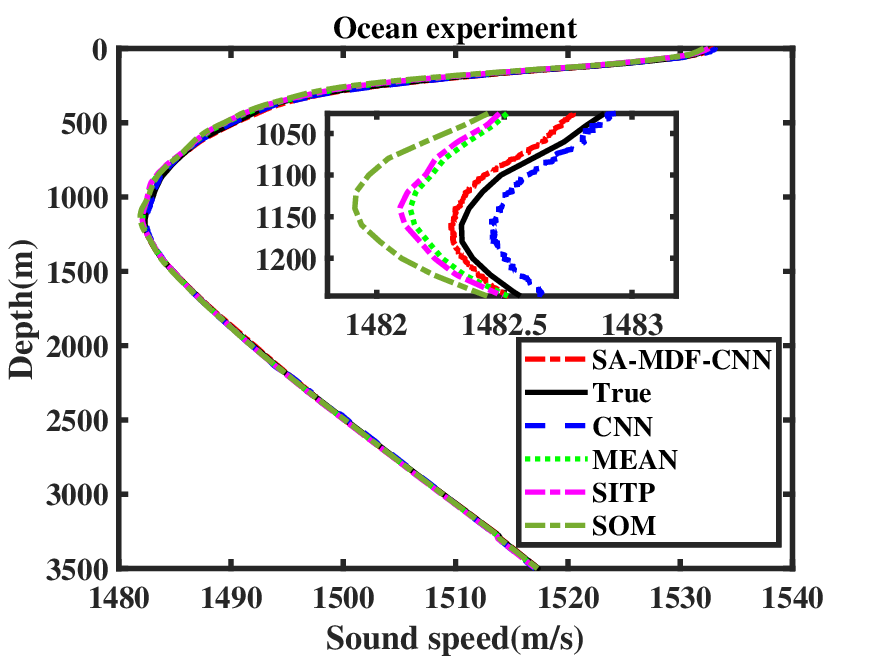}}
	\subfloat[]{\includegraphics[width = 0.24\textwidth]{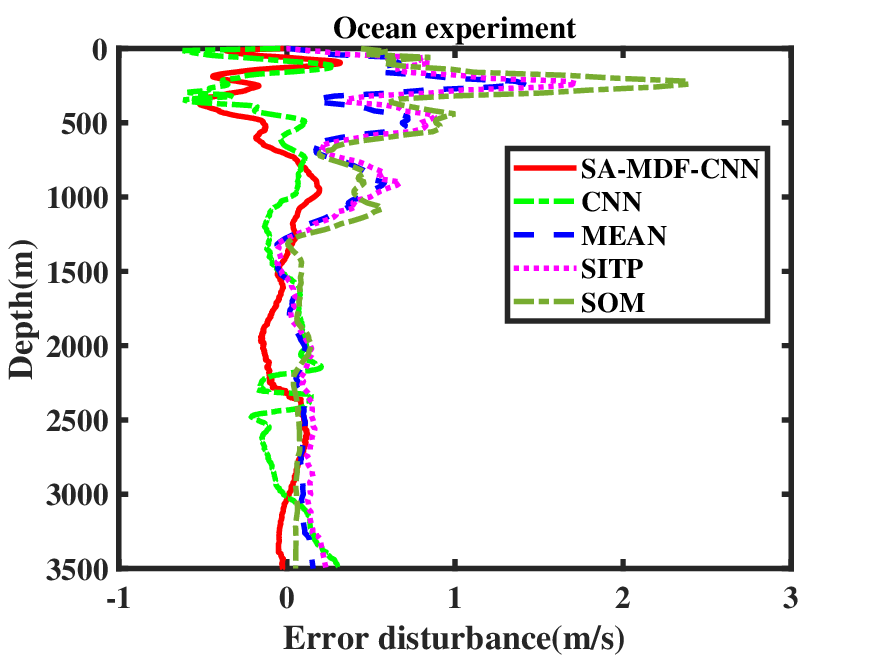}}
	\caption{Evaluation of real-time SSP estimation outcomes utilizing diverse algorithms on oceanographic experimental data, where (a) shows the SSP curves, and (b) shows the error distributions.}
	\label{fig19}
\end{figure}


In order to verify the practicality of the SA-MDF-CNN model, a systematic abyssal marine research initiative was executed within the South China Sea basin throughout April 2023. As illustrated in figure~\ref{fig18} (a), various instruments such as CTD and XCTD were employed at diverse marine locations, yielding a total of 14 distinct samples. The depth of sampling reached 3500 meters, with intervals set at every 100 meters. It is evident that the selected location for this experiment does not conform to the rigid 3x3 grid pattern, thus allowing for a more rigorous evaluation of the proposed model. The input data for the model were extracted from the fusion dataset corresponding to the blue coordinates in figure~\ref{fig18} (b), while the real-time SSP associated with the red pentagram served as the model's predictive output.  

\begin{figure*}[!h]
    \centering
    \subfloat[]{\includegraphics[width = 0.24\textwidth]{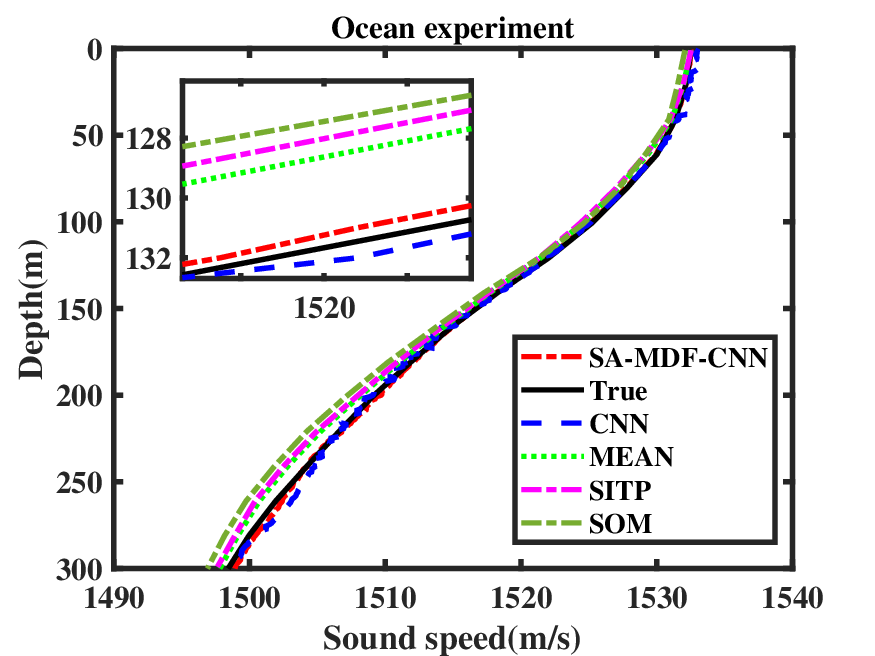}}
    \subfloat[]{\includegraphics[width = 0.24\textwidth]{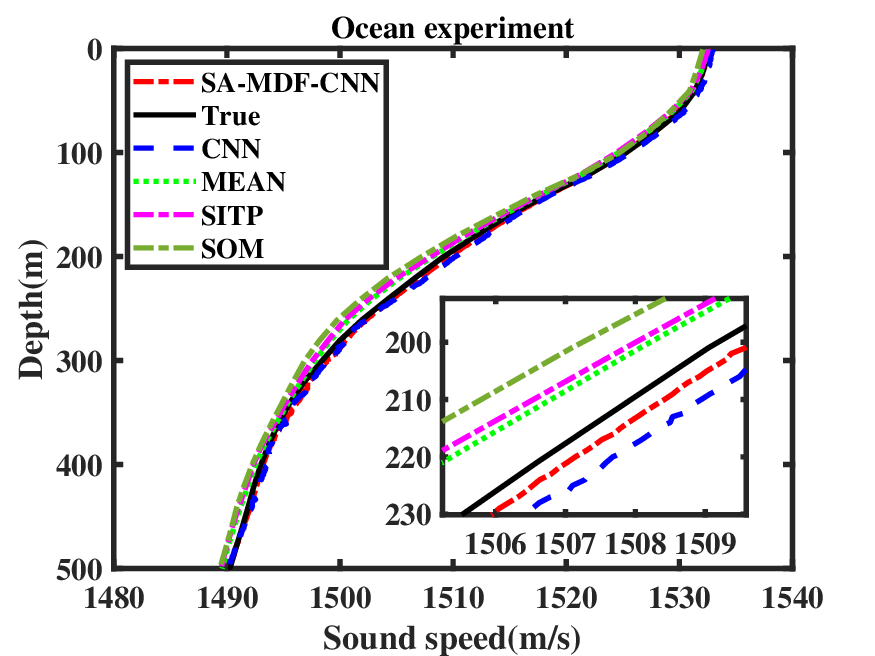}}
    \subfloat[]{\includegraphics[width = 0.24\textwidth]{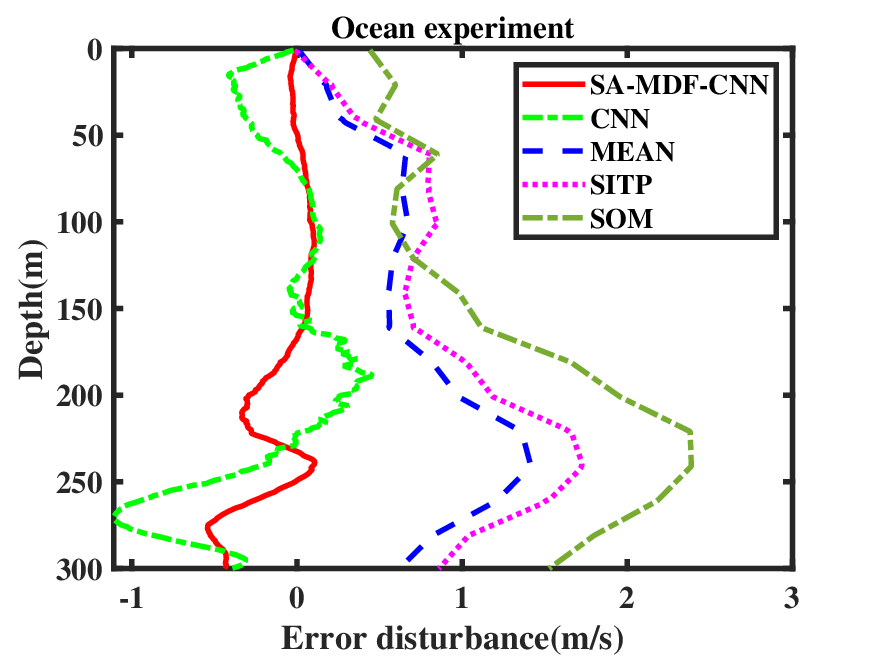}}
    \subfloat[]{\includegraphics[width = 0.24\textwidth]{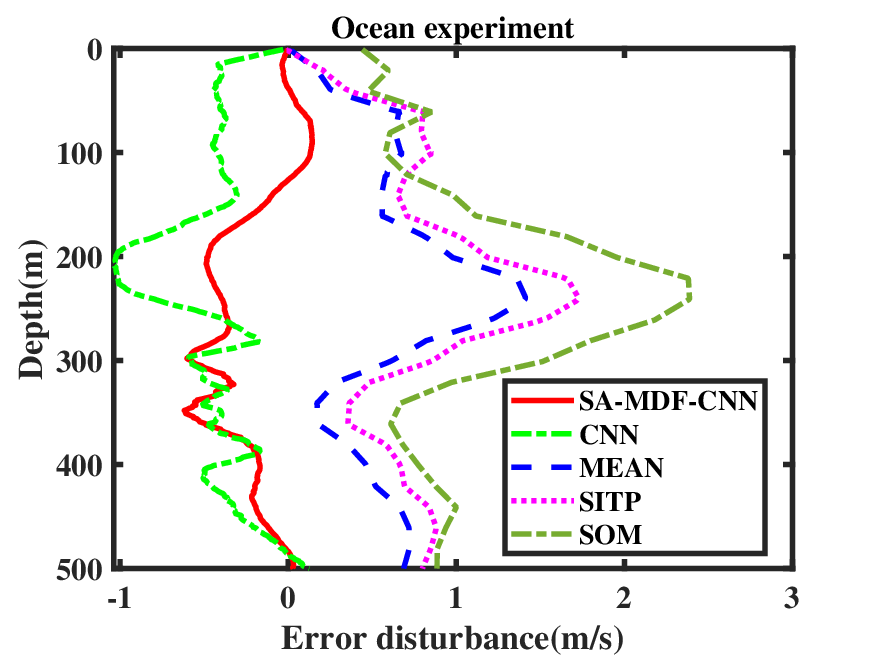}}
    \caption{Comparison of real-time estimation SSP results of different algorithms in shallow sea environment with different depths in ocean experiment, where (a) and (c) are SSP curves of 300 meters depth, (b) and (d) are error distributions of 500 meters depth.}
    \label{fig20}
\end{figure*}

\begin{figure*}[!htbp]
	\centering
	\includegraphics[width=\textwidth]{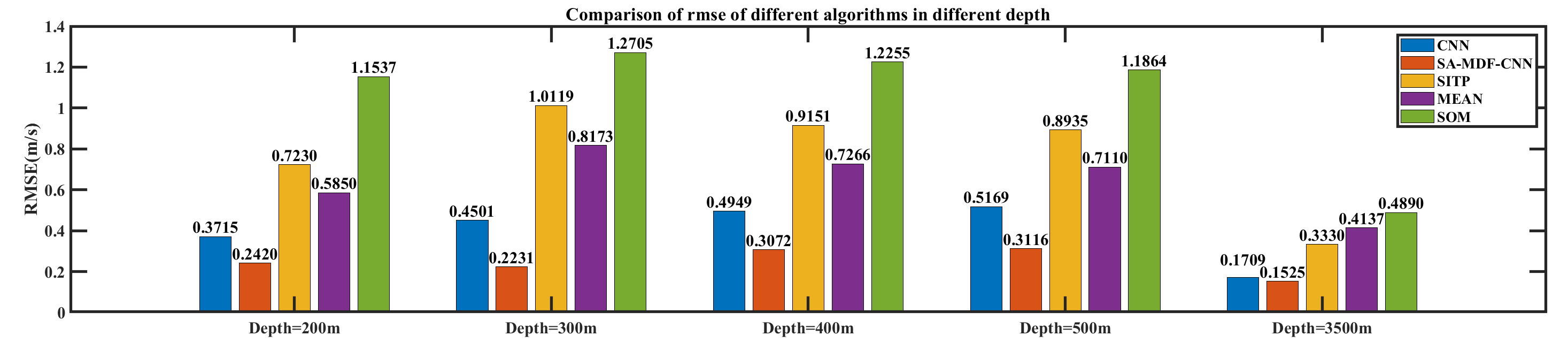}
	\caption{RMSE comparison of real-time estimated SSPs by different algorithms of ocean experiments.}
	\label{fig21}
\end{figure*}

To evaluate the effectiveness of the model in estimating the distribution of sound velocity at full-ocean depth, the estimated SSP of SA-MDF-CNN is compared with other algorithms in figure~\ref{fig19}. It is evident that the proposed model exhibits lower RMSE and more stable error disturbance compared to traditional algorithms and CNN. To further test the sound velocity estimation performance in shallow waters with significant changes in sound velocity, estimation of sound velocity at 300, 500 depth meters by SA-MDF-CNN are compared with other methods in figure~\ref{fig20}. The experimental outcomes demonstrate that the introduced methodology exhibits reduced error fluctuations and sustains notable precision benefits relative to alternative approaches in shallow marine settings. This comparative analysis highlights the robustness and superior performance of the proposed technique under such environmental conditions. Figure~\ref{fig21} gives a more detailed comparison of the RMSE results of different algorithms, clearly showing that the proposed model has significant advantages at different depths.

\section{Conclusion}
To construct a real-time sound velocity field and eliminate the need for underwater onsite data measurement operations, a SA-MDF-CNN model driven by multimodal data fusion is proposed in this paper, which fuses historical SSP features, real-time remote sensing SST, latitude and longitude coordinates. The fusion data is constructed by gridding, and the real-time SSP is retrieved based on the real-time remote sensing SST. The effectiveness of our proposed model was tested in the case of non-standardized grid through ocean experiments. The empirical findings demonstrate that the developed framework exhibits dual applicability, functioning effectively for  real-time SSP reconstruction in non-uniform grid environments while maintaining effectiveness across diverse shallow-water bathymetric scenarios.

\section*{Acknowledgments}
The authors acknowledge the historical SSP data support from the China Argo Real-time Data Center, (https://www.argo.org.cn/, latest access:December 20, 2024). The remote sensing SST data can be acquired at the National
Oceanic and Atmospheric Administration (https://www.commerce.gov/, latest access: December 25, 2024).

\bibliographystyle{IEEEtran}
\bibliography{IEEEabrv,draft}

\vfill

\end{document}